# ADMINISTRATIVE LAW'S FOURTH SETTLEMENT: AI AND THE SCRUTABLE STATE

*Nicholas Caputo*[*]

ABSTRACT

Since 1887, administrative law has confronted a problem of institutional cognition. Expert agencies are needed to govern systems of increasing technological complexity, but expertise makes agency decisions difficult for courts, Congress, and the public to understand and oversee. Administrative law has responded to this "capability-accountability trap" by requiring records, reason-giving, and transparency, drawn together through procedural review. These devices have preserved legality, but procedural requirements have piled up, making government both less comprehensible and less effective.

This Article offers a new account of the Supreme Court's recent administrative law retrenchment, rooted in these fundamental problems of institutional structure and information-processing. From *Loper Bright* through *Trump v. Slaughter*, the Court has reallocated authority to entities it regards as comprehensible and attributable. It is attempting to restore accountability by making government "scrutable," comprehensible to its overseers and the public, but in doing so it is sacrificing capability and undermining the effectiveness of administration.

AI offers a different path. Deployed correctly, AI could help make government both more effective and more transparent, translating technical complexity into accessible terms, surfacing the assumptions that matter for oversight, and enabling substantive verification of agency reasoning. This technical integration must be accompanied by an updated administrative law, built around a Model and System Dossier (documenting model purpose, evaluation, monitoring, and versioning) that extends the administrative record to AI decision-making; a material-model-change trigger specifying when AI updates require new process; and a deference to audit standard that rewards agencies for auditable evaluation of their AI tools. The result is a framework for what this Article calls the "Fourth Settlement," administrative law that escapes the capability-accountability trap by preserving capability while restoring comprehensible oversight of administration.

---

[*] Assistant Professor, Johns Hopkins School of Government and Policy. Law & AI Lead, Oxford Martin AI Governance Initiative, University of Oxford. J.D., Harvard Law School. Thanks to Jack Boeglin, Molly Brady, Glenn Cohen, Noah Feldman, Gillian Hadfield, Noam Kolt, Alex Mechanick, Peter Salib, Ben Rolsma, and Justin Walker for their insightful comments. Thanks to Ryan Copus, Guha Krishnamurthy, Alex Platt, and the faculty of the Johns Hopkins University School of Government and Policy for stress testing these arguments.

*Administrative theory is peculiarly the theory of intended and bounded rationality—of the behavior of human beings who satisfice because they have not the wits to maximize.*

– Herbert Simon, *Administrative Behavior* (1957)[1]

## INTRODUCTION

On June 29, 2026, the Supreme Court overruled *Humphrey's Executor*,[2] ending the ninety-year settlement that had allowed Congress to insulate expert agencies from presidential removal. "If anything more is left of Humphrey's," the Chief Justice wrote, "we overrule it."[3] The decision in *Trump v. Slaughter* marked two remarkable years of doctrinal upheaval in administrative law. In 2024, *Loper Bright* withdrew judicial deference to agency interpretations of law.[4] The major questions doctrine continued its resurgence, limiting agency power over issues of broad public significance.[5] *Jarkesy* stripped agencies of in-house adjudication of civil penalties.[6] And alongside the Court, the Executive has pursued its own robust program of governmental simplification. Reductions in force, agencies reduced to statutory minima, and career officials reclassified for at-will removal are only some of the steps taken.[7] American administration is being made smaller, simpler, and more subordinate than at any point since 1935.

---

[1] HERBERT SIMON, ADMINISTRATIVE BEHAVIOR: A STUDY OF DECISION-MAKING PROCESSES IN ADMINISTRATIVE ORGANIZATION (2d ed. 1957).

[2] 295 U.S. 602 (1935).

[3] Trump v. Slaughter, 609 U.S. ____, slip op. at 21 (2026).

In the same two years, those same agencies reported more than 3,600 uses of artificial intelligence (AI), several hundred designated "high-impact," under executive direction to adopt the technology faster.[8] The federal government is reducing its expert human workforce and the institutional independence and discretion associated with it just as it begins adopting what is likely the most powerful cognitive technology ever built. These two developments are usually told as separate stories, one of law and one of technology. This Article argues that they are one story, and that it began in 1887.

That year, Congress created the Interstate Commerce Commission (ICC). The ICC was the first modern federal regulatory agency, staffed by experts, armed with tools of statistics and telegraphy,

---

[4] Loper Bright Enters. v. Raimondo, 603 U.S. 369, 391–94, 412–13 (2024).

[5] *See* West Virginia v. EPA, 597 U.S. 697, 721–24 (2022).

[6] SEC v. Jarkesy, 603 U.S. 109 (2024).

[7] *See* Exec. Order No. 14,210, § 3(c), 90 Fed. Reg. 9,669, 9,670 (Feb. 14, 2025) (directing agency heads to prepare large-scale reductions in force); Exec. Order No. 14,238, § 2(a), 90 Fed. Reg. 13,043, 13,043 (Mar. 20, 2025) (ordering the elimination, to the maximum extent consistent with law, of the nonstatutory components and functions of seven governmental entities and the reduction of their statutory functions and personnel to the minimum required by law); *Improving Performance, Accountability and Responsiveness in the Civil Service*, 91 Fed. Reg. 5,580, 5,580, 5,585 (Feb. 6, 2026) (establishing Schedule Policy/Career positions as at-will positions excepted from adverse-action procedures and appeals), implemented by Exec. Order No. 14,410, 91 Fed. Reg. 34,893, 34,893–94 (June 10, 2026).

[8] Office of Mgmt. & Budget, Exec. Office of the President, *2025 Federal Agency Artificial Intelligence Use Case Inventory* (updated Apr. 13, 2026) (reporting 3,611 AI use cases, including 445 high-impact); Office of Mgmt. & Budget, Exec. Office of the President, Memorandum M-25-21, *Accelerating Federal Use of AI Through Innovation, Governance, and Public Trust* 1–2 (Apr. 3, 2025) (directing agencies to "accelerate the Federal use of AI").

and insulated from the politics of the moment.[9] It was built to master a transformative new technology, the railroads, that had overwhelmed existing governance institutions.[10] The ICC embodied the first modern institutional response to a structural dilemma that has shaped administrative governance ever since, what this Article calls the "capability-accountability trap."[11] To solve the novel problems presented by novel technologies, government must become more scientifically and technologically sophisticated. But this sophistication and the resulting complexity threaten to undermine democratic control because oversight bodies, namely the courts, Congress, and the public they serve, cannot easily understand how scientific government works and so cannot ensure that it is acting in the public interest.[12] The railroad ratemakings of the late

---

[9] THOMAS K. MCCRAW, PROPHETS OF REGULATION: CHARLES FRANCIS ADAMS, LOUIS D. BRANDEIS, JAMES M. LANDIS, ALFRED E. KAHN 61–62 (1984). Professor Jerry L. Mashaw has argued that administrative law predates the ICC and structured earlier forms of American government. *See generally* JERRY L. MASHAW, CREATING THE ADMINISTRATIVE CONSTITUTION: THE LOST ONE HUNDRED YEARS OF AMERICAN ADMINISTRATIVE LAW (2012). However, those earlier forms were much narrower than the modern administrative state, and today's administrative law doctrine looks to the debates over the ICC as its starting point. *See* Susan E. Dudley, *Milestones in the Evolution of the Administrative State*, 150 DAEDALUS 33, 34 (2021) (describing the ICC as the first modern regulatory agency).

[10] McCraw, *supra* note 9, at 61–62; STEPHEN SKOWRONEK, BUILDING A NEW AMERICAN STATE: THE EXPANSION OF NATIONAL ADMINISTRATIVE CAPACITIES, 1877-1920 4–5 (1982) ("Generally speaking, the expansion of national administrative capacities in America around the turn of the century was a response to industrialism.").

[11] My framing builds on Professor Adrian Vermeule's "deference dilemma," describing how the background conditions of complexity in administration necessitate broad and vague delegations of power to agencies at the same time that the need for legitimating judicial review pushes back against such delegations. Adrian Vermeule, *The Deference Dilemma*, 31 GEO. MASON L. REV. 619, 619–20 (2024). However, whereas Professor Vermeule focuses on the implications of his dilemma for delegation of interpretive authority, I argue more broadly that the tension between administrative capability and accountability structures the whole of administrative law and also draw the connection to how new technologies change this picture.

[12] *See generally* MAX WEBER, ECONOMY AND SOCIETY: A NEW TRANSLATION 350–54 (Keith Tribe ed., trans., Harv. U. Press 2019) (1921) (describing the growth of technically trained bureaucracies as a functional response to new technology and to complex social and economic problems, and warning that rationalization and expert administration may be necessary but also threaten democratic self-rule by concentrating authority in domains inaccessible to lay understanding); JAMES M. LANDIS, THE ADMINISTRATIVE PROCESS 23–30 (1938) (arguing that modern industrial conditions require governance by scientifically trained experts while acknowledging the resulting tension with democratic accountability); HERBERT A. SIMON, ADMINISTRATIVE BEHAVIOR 46–47, 65–67, 87–118 (4th ed. 1997) (explaining that bounded rationality necessitates reliance on expert specialization in complex decision environments,

nineteenth century were delegated to expert commissions of administrators because judges and legislatures could not act with sufficient speed and sophistication to handle them.[13] Further technological and scientific progress added difficulty: in the deliberations over *Chevron*, with its technical debate over scientific meaning and emissions modeling, Justice John Paul Stevens admitted that "[w]hen I get so confused, I go with the agency."[14] If oversight is a criterion of the legitimacy of administration,[15] inscrutability undermines it.

Here, "capability" refers to the government's ability to process information, coordinate action, and implement solutions at the scale and sophistication that the problems facing it demand. "Accountability" encompasses both democratic control, ensuring that government responds to the public will through direction and oversight, and legality, maintaining rule-of-law constraints on arbitrary power. Both conditions are necessary for modern government, but historically they have traded off[16] because increasing capability has required increasing complexity and so incomprehensibility, while increasing accountability has required reducing those things.

---

with implications for democratic control over value judgments in decision-making); SHEILA JASANOFF, THE FIFTH BRANCH: SCIENCE ADVISERS AS POLICYMAKERS 1–32 (1990).

[13] *See* GABRIEL KOLKO, RAILROADS AND REGULATION, 1877–1916, 57–150 (1976); McCraw, *supra* note 9, at 62.

[14] Mark Sherman, *High Court Climate Case Looks at EPA's Power*, HUFF POST (Feb. 23, 2014, 3:59 PM), https://perma.cc/SQ6K-5ANA.

[15] LOUIS L. JAFFE, JUDICIAL CONTROL OF ADMINISTRATIVE ACTION 320 (1965) ("The availability of judicial review is the necessary condition, psychologically if not logically, of a system of administrative power which purports to be legitimate, or legally valid.").

[16] *See* Nicholas R. Bednar, *Presidential Control and Administrative Capacity*, 77 STAN. L. REV. 823, 825–32 (2025) (discussing the empirical correlation between independence and capability); Elena Kagan, *Presidential Administration*, 114 HARV. L. REV. 2245, 2352–58 (2001) (framing "the apparent tradeoff between politics and expertise" in administrative decisionmaking as a central problem for institutional design and oversight); Matthew C. Stephenson, *Optimal Political Control of the Bureaucracy*, 107 MICH. L. REV. 53, 53–56 (2008) (taking as a starting point the conventional view that greater bureaucratic insulation can improve expert decisionmaking but "reduces the agency's responsiveness to the preferences of political majorities"); Mathew D. McCubbins, Roger G. Noll, & Barry R. Weingast, *Administrative Procedures as Instruments of Political Control*, 3 J.L. ECON. & ORG. 243, 244–45, 259–61 (1987) (arguing that delegation generates principal-agent problems because agencies may become "more expert" than elected principals; describing administrative procedures as devices to reduce informational asymmetries and reassert democratic/political control).

Capability and accountability do not trade off directly. Their relationship is mediated by *scrutability*, the cognitive tractability of administrative action for those who must oversee, contest, and participate in it.[17] Scrutability is not itself accountability, but a precondition for it, because government action that cannot be understood or attributed cannot be meaningfully reviewed, contested, or sanctioned. Scrutability includes both decisional scrutability, whether outsiders can understand the reasons, evidence, assumptions, and tradeoffs that went into a specific decision, and responsibility scrutability, whether they can identify the official or institution responsible for a given action.

Across the history of public administration, as administrative problems became more complex, substantive oversight became less feasible. Administrative law responded through three recurring devices: reallocating authority among agencies and their overseers; requiring reasoned decision-making on an administrative record; and forcing disclosure of the information necessary for review. Allocation clarified who was responsible; records and disclosure made agency action more inspectable. Together, these devices made administration reviewable without requiring generalist overseers to replicate agency expertise. Increasingly, they did so by substituting procedural proxies for substantive review. It is no accident that the Administrative Procedure Act remains the centerpiece of administrative law, and where courts do substantively intervene, it is often to decide questions of plausibility rather than actual correctness.[18] Across three historical settlements of administrative law—the railroads and the expert commission, the Depression and the Administrative Procedure Act, and complex science and *Chevron*—these devices preserved a form of accountability while enabling government to retain the capability its tasks demanded. Each settlement made particular agency decisions more reviewable. Cumulatively, however, the resulting layers made the administrative system itself harder to understand and use. Over time, the accumulated architecture became a source of complexity in its own right: the "scrutability tax," paid in delay, cost, defensive documentation, and increasing distance between administration and the public. Take, for example, the recent revision of the National Ambient Air Quality Standard for $PM_{2.5}$. It took three years to change one number, required the processing of 700,000 comments, and produced a 200-page preamble in what was intended to be a "concise general statement of [. . .] basis and purpose."[19] Today, inscrutability is both substantive and procedural, as agencies and their overseers now must understand and resolve difficult scientific disputes while also following accumulated procedural requirements made more burdensome through litigation by affected parties. The result is government that feels neither effective nor legitimate.

[17] I use "scrutability" as something like the inverse of James C. Scott's "legibility." Where "legibility" means the comprehensibility of the world to regulators, I mean the comprehensibility of regulators to the world. *See* JAMES C. SCOTT, SEEING LIKE A STATE 2–3 (1998). *See also* Andrew D. Selbst & Solon Barocas, *The Intuitive Appeal of Explainable Machines*, 87 FORDHAM L. REV. 1085, 1094 (2018) ( "inscrutability" as "a situation in which the rules that govern decision-making are so complex, numerous, and interdependent that they defy practical inspection and resist comprehension"); Katherine J. Strandburg, *Rulemaking and Inscrutable Automated Decision Tools*, 119 COLUM. L. REV. 1851, 1851–54 (2019).

[18] *See* Citizens to Preserve Overton Park v. Volpe, 401 U.S. 402, 420 (1971) (asking if the agency made a "clear error in judgment").

This growing inscrutability has now produced a backlash, but one the historical pattern does not predict, because it arrives without the surge in governmental capability that preceded the prior unsettlements. The Court's recent administrative law cases can be understood as efforts to restore scrutability by relocating authority into institutions and forms it regards as comprehensible. *Loper Bright* reclaims a degree of legal interpretation for the courts.[20] The major questions doctrine requires Congress itself to authorize major policy choices in clear and attributable terms.[21] *Jarkesy* shifts certain civil-penalty adjudications into Article III courts and jury proceedings.[22] And *Slaughter* concentrates more responsibility in the President, simplifying public attribution.[23] These cases each shorten or clarify the chain by which official action is made, reviewed, or attributed. This scrutability framework adds explanatory power beyond anti-administrativism or partisanship because it accounts for the Court's continued deference in cases where deference reduces complexity, for example by reducing the procedural burden of environmental analysis in

[19] Reconsideration of the National Ambient Air Quality Standards for Particulate Matter, 89 Fed. Reg. 16,202 (Mar. 6, 2024) (to be codified at 40 C.F.R. pts. 50, 53, 58). For the statutory reference, see 5 U.S.C. § 553(c) (2018).

[20] 603 U.S. at 391–94, 412–13.

[21] 597 U.S. at 721–24.

[22] Jarkesy, slip op. at 20–21.

[23] Slaughter, slip op. at 35–36.

*Seven County*[24] and by declining to revive the nondelegation doctrine in *Consumers' Research*.[25] And the Court's diagnosis is broadly correct: the complexity crisis is real, and accountability has eroded. However, the Court's remedy is not. Stuck in the capability-accountability trap, its moves have sought to resolve the accountability deficit by deepening the capability deficit, dismantling expert governance precisely as new risks from climate change, pandemic risk, and advanced AI demand more of it.

But the path back to *Chevron*-era deference and ever more ex post procedural review does not solve the problem. Escaping the trap requires changing the cost of comprehension of government action. Courts, Congress, and the public need tools that let them oversee complex administration without forcing agencies to become simple.

Artificial intelligence could provide an opportunity for precisely this reconfiguration. Earlier administrative technologies sometimes improved legibility by generating standardized records and new means of review, but the complexity they enabled still had to be processed by scarce human experts. General-purpose AI may operate on administrative complexity itself, by translating technical material, surfacing the assumptions on which agency conclusions depend, and allowing courts, Congress, and the public to interrogate large records at much lower marginal cost. Agencies could use the technology to process masses of claims and comments, while their overseers use it to examine the resulting decisions, giving courts and Congress better means to test the assumptions, evidence, and tradeoffs reflected in agency reasoning, rather than merely confirming that a paper trail exists.

To govern this AI-assisted administration, this Article proposes three core doctrinal innovations that would provide the foundation for a new settlement in administrative law. First, a "Model and System Dossier" extends the administrative record to AI, documenting the purpose, evaluation, monitoring, and versioning of each AI system used in an administrative action. Second, the "material-model-change trigger" specifies when updating a system amounts to amending the rule it applies, and so requires process. And third, "deference to audit" reorients judicial review away from deference to agency expertise and toward verification that agencies have audited their uses of AI according to set standards of quality, fairness, and accuracy. This shift plugs judicial oversight of government uses of AI into the advancing science of AI auditing and evaluation, ensuring that oversight will keep pace through its own technological innovations. And while it operates on a procedural substrate to enable judicial review, deference to audit would likely reduce the overall procedural burden faced by agencies and overseers because system-level evaluations can often be reused across many decisions, allowing oversight costs to grow more slowly than the volume and complexity of agency action. The basic architecture described herein could be implemented with current systems for retrieval, summarization, classification, and structured evaluation all grounded in the administrative record, though implementation would require domain-specific testing, monitoring, and legal safeguards.

[24] Seven County Infrastructure Coalition v. Eagle County, 605 U.S. ____, slip op. at 2–3 (2025).

[25] FCC v. Consumers' Research, 606 U.S. ____, slip op. at 12–13 (2025).

Separate from this implementable architecture is a more speculative possibility, that of restoring more meaningful substantive review of agency actions. Advances in technical alignment,[26] interpretability,[27] and scalable oversight[28] may eventually permit forms of substantive verification impossible with human administrators. If those methods can reliably show that a system pursued authorized objectives and relied only on legally-relevant considerations, administrative law could reduce its dependence on procedural proxies and bring something like substantive review back in.

This Article makes two main contributions: First, it offers a structural account of the recent administrative retrenchment that also explains its exceptions, locating the Court's project within a 140-year dynamic of capability and accountability rather than focusing on the politics of the moment. Second, the Article makes an affirmative case for AI as both administrative and oversight infrastructure. History teaches that government will have to integrate AI to solve the many new problems we are confronting today, and AI's distinctive promise is that it may allow the cost of understanding and verifying administrative action to grow more slowly than the complexity of the action itself. This argument adds a constructive account of the uses of AI in government to the literature of algorithmic governance, which has largely focused on how to constrain the risks of governmental AI, and helps update it for the frontier AI era.

---

[26] *See* Long Ouyang et al., *Training Language Models to Follow Instructions with Human Feedback*, arXiv:2203.02155 (Mar. 4, 2022) (introducing RLHF for instruction-following and reporting improvements in human preference, truthfulness, and toxicity); Yuntao Bai et al., *Constitutional AI: Harmlessness from AI Feedback*, arXiv:2212.08073 (Dec. 15, 2022) (principle-based training with self-critique and "RL from AI feedback" (RLAIF) to scale supervision and improve harmlessness).

[27] *See* Finale Doshi-Velez & Been Kim, *Towards A Rigorous Science of Interpretable Machine Learning*, arXiv:1702.08608 (Mar. 2, 2017).

[28] *See* Samuel R. Bowman et al., *Measuring Progress on Scalable Oversight for Large Language Models*, arXiv:2211.03540 (Nov. 11, 2022).

The Article proceeds as follows. Part I traces the history of administrative law through three prior settlements, showing how doctrine repeatedly mediated the coevolution of governance and technology, and how each settlement added procedural complexity that would strain the next. Part II explains the current crisis of administration. It argues that the problem is not capability outrunning accountability, but rather both eroding under accumulated scientific complexity and procedure, and describes the judicial and executive dismantlement that erosion has provoked. Part III proposes a Fourth Settlement. It lays out a framework for AI as scrutability infrastructure for participation and oversight, an accountability architecture for AI-assisted administration, and the path from procedural to substantive accountability.

Whether AI develops along the trajectory that would make this settlement possible, whether political conditions will permit its adoption, and whether agencies and courts could implement it well are all uncertain. But the current moment demands more than diagnosis of administrative law's failures. It demands a constructive vision of what legitimate and effective administration could look like in this era of technological transformation. The previous settlements were not inevitable. They were built by reformers who saw that new technological tools could mediate the tension between capability and accountability in government in new ways. This Article aims to do the same by describing one possible settlement for the AI era. AI is already changing both the problems government must address and the tools available to address them. The question is whether democratic societies can use those tools to build government sophisticated enough for contemporary challenges and scrutable enough for democratic control.

## I. A TECHNOLOGICAL HISTORY OF ADMINISTRATIVE LAW

The American administrative state has evolved through recurring cycles of technological disruption and legal adaptation. This Part traces three such cycles—the railroads, the New Deal, and computerized risk regulation—to show how technological shocks increased administrative capability while pushing accountability toward procedural instead of substantive review. Understanding this pattern is essential for two reasons. First, it reveals that our current administrative framework is not the product of abstract constitutional theory but rather accumulated responses to concrete technological problems. Second, it exposes why each doctrinal settlement has proven temporary: new technologies have presented new problems that force adjustments in the administrative state and add to the complexity of administration and its oversight.

Fully developing each historical case, particularly the political contestation that drives and shapes lawmaking,[29] exceeds the scope of this Article, but the pattern of capability crisis, government complexification, accountability gap, and doctrinal settlement clearly emerges across them. Importantly, each step in the cycle also represents an increase in the overall complexity of administration as scientific problems have demanded ever more sophisticated government, increasing the burdens on the administrative settlements and distancing administration from democratic control by introducing more procedural intermediation.

---

[29] The ICC emerged not mechanically from railroad complexity but from decades of agrarian mobilization, Granger legislation, and intraparty conflict over the terms of federal regulation. *See* ELIZABETH SANDERS, ROOTS OF REFORM: FARMERS, WORKERS, AND THE AMERICAN STATE, 1877–1917 179–213 (1999). The New Deal agencies reflected coalition politics and constitutional brinkmanship as much as functional demands. *See* STEPHEN SKOWRONEK, THE POLITICS PRESIDENTS MAKE: LEADERSHIP FROM JOHN ADAMS TO BILL CLINTON 287–324 (1997). And the *Chevron* settlement was shaped by deregulatory politics and judicial ideology, not only by the cognitive limits of courts. *See* Thomas W. Merrill, *The Story of* Chevron: *The Making of an Accidental Landmark*, 66 ADMIN. L. REV. 253, 275–82 (2014). This Article foregrounds the technological dimension not because politics was secondary but because the relationship between technological change and institutional adaptation has received less systematic attention in administrative law scholarship, aiming for "pattern identification" rather than full explanation. KAREN ORREN & STEPHEN SKOWRONEK, THE SEARCH FOR AMERICAN POLITICAL DEVELOPMENT 7–8 (2004).

### *A. Railroads, the Telegraph, and the Birth of Expert Administration (1850–1920)*

On August 26, 1871, a speeding locomotive plowed into a stationary passenger train at Revere station near Boston, Massachusetts.[30] Twenty-nine passengers died in the wreckage. The disaster emerged from a cascade of failures exemplifying the governance crisis railroads had created. Although the railroad was found liable for negligence, an investigation found that "the corporation had broken no existing law."[31] Traditional legal mechanisms struggled to address the systemic complexity of railroad operations.[32] The tools of pre-industrial governance had become inadequate for rapidly developing industrial problems.[33]

This inadequacy reflected a fundamental capability crisis. Between 1850 and 1890, railroad track in operation expanded from 9,000 to 166,000 miles,[34] while the telegraph enabled instantaneous communication across the country.[35] Together, these technologies knit local markets into national ones and created new problems and forms of economic power that existing institutions could not comprehend or control, as the railroads demonstrate.[36] Passengers, passers-by, and employees were killed and injured in large numbers by speeding trains, and derailment catastrophes like the one at Revere occurred with some frequency.[37] Rate discrimination by the railroads harmed travelers and producers, but state legislatures and courts lacked the sophistication to determine what setting "just and reasonable"[38] rates for a particular railroad might mean. Rate decisions crossed jurisdictional lines, such that a railroad's intrastate rates in Illinois could devastate Iowa farmers shipping to Chicago, but state governments lacked the authority to regulate outside their jurisdiction after the Supreme Court's decision in the *Wabash* case.[39] Discriminatory pricing that favored long-distance shippers over local ones could shift patterns of economic development.[40] Aggressive rate regulation might protect local shippers but drive railroads to recoup losses through higher interstate rates beyond the state's control, while lenient regulation invited exploitation of captive local markets to subsidize competitive long-haul routes.

[30] McCraw, *supra* note 9, at 26–28.

[31] *Id.*

[32] Bruce Wyman, *The Rise of the Interstate Commerce Commission*, 24 YALE L.J. 529, 531 (1915).

[33] The pace of technological change began to increase during the nineteenth century as the technologies of the Industrial Revolution disseminated around the world. *See* JOEL MOKYR, THE LEVER OF RICHES 82–83 (1990).

[34] U.S. BUREAU OF THE CENSUS, HISTORICAL STATISTICS OF THE UNITED STATES: COLONIAL TIMES TO 1957 427–28, ser. Q 15–22 (U.S. Dep't of Commerce 1960) (reporting U.S. railroad mileage over time).

[35] ALFRED D. CHANDLER, JR., THE VISIBLE HAND: THE MANAGERIAL REVOLUTION IN AMERICAN BUSINESS 79–107 (1977).

[36] *Id.*

[37] McCraw, *supra* note 9, at 25–26.

This governmental capability crisis eventually precipitated a response, in the form of the expert commission. Massachusetts created the first expert railroad commission in 1869, staffed by specialists who could comprehend railroad operations using the industry's own tools of statistical analysis, standardized accounting, and telegraph communication.[41] As Professor Thomas McCraw writes, this "agency began life as an institutional adaptation called into being because, after a generation of uninterrupted growth, the railroad system had become too much for the old legal machinery to handle."[42] In subsequent years, state railroad commissions sprang up in other states with varying degrees of legal capability.[43]

The state commissions set an important precedent for the creation of the ICC in 1887, which sought to provide a national solution to the problems of national railroad governance.[44] Strengthened over time through a series of laws expanding its powers and jurisdiction,[45] the ICC was the first federal regulatory agency with legal protections guaranteeing its independence and quasi-legislative and quasi-judicial powers to set rates and adjudicate cases, require standardized accounting practices, and adjudicate disputes across state lines.[46] It compelled testimony, examined books, and mandated uniform reporting systems that allowed for the first truly national picture of railroad operations. Yet the ICC faced immediate challenges that revealed the persistent capability gap that railroad governance still faced. In particular, its initial powers were severely limited by judicial review often favoring the railroads, characteristic of the broader attitude of *laissez-faire* government that the administrative state would struggle with into the 1930s.[47]

---

[38] Interstate Commerce Act of 1887, ch. 104, 24 Stat. 379 § 1 (Feb. 4, 1887) (repealed and recodified as 49 U.S.C. § 10101 et seq. (Supp. III 1980)).

[39] 118 U.S. 557, 564–66, 577–79 (1886).

[40] *See* Felix L. Barton, *Economic Effects of Discriminatory Freight Rates*, 12 L. & CONTEMP. PROBS. 507, 520–31 (1947).

[41] McCraw, *supra* note 9, at 19–21.

[42] *Id.* at 20.

[43] *Id.* at 57.

[44] *Id.* at 61–62.

[45] *Id.* at 62–63.

The emergent expert commission form, with its specialist staffs and partial insulation from politics, increased government capability but threatened accountability. Should unelected commissioners make rules and adjudicate cases so deeply affecting American life? When the ICC set rates, affected shippers had little recourse if commissioners misunderstood market conditions because judges could not substantively evaluate the agency's economic analysis, legislators met too infrequently to effectively intervene, and voters could not remove appointed commissioners.

The legal cases immediately brought against the ICC upon its founding illustrate these worries and demonstrate how courts struggled with balancing the need for expert government to handle the railroads with legitimacy and accountability. Early cases went badly for the ICC, "reduc[ing] it, by the late 1890s, to a mere collector of data."[48] *ICC v. Cincinnati, N.O. & Texas Pacific Ry. Co.*,[49] decided in 1897, is characteristic: the ICC, set up to resolve the problem of railroad rate discrimination, sought to do so in part by setting prospective national railroad rates pursuant to what it understood to be a grant of authority from Congress in the Interstate Commerce Act of 1887 to set "reasonable rates."[50] However, the Supreme Court held that the power to set rates was a legislative power that Congress could only constitutionally delegate through explicit language, which it had not done.[51] As such, the ICC had acted in excess of its authority in setting rates and enjoining the plaintiff railroad to follow them,[52] despite these actions being taken to address the very regulatory problem that it had been established to resolve.

[46] Interstate Commerce Act of 1887, ch. 104, § 11, 24 Stat. 379, 383–84 (creating a five-member commission; fixed terms; bipartisan limitation; and for-cause removal).

[47] Robert L. Rabin, *Federal Regulation in Historical Perspective*, 38 STAN. L. REV. 1189, 1208–35 (1986) (situating the ICC within Populist and Progressive conflicts, particularly around the Supreme Court).

[48] McCraw, *supra* note 9, at 62.

[49] 167 U.S. 479 (1897).

[50] 167 U.S. at 485–88.

But the problems of the railroads did not go away, and between 1897 and 1920 a doctrinal settlement emerged that managed, though did not resolve, these tensions.[53] This doctrinal settlement established the three characteristic tools through which administrative law handles the capability-accountability trap. First, *allocation of authority*: Congress responded to *ICC v. Cincinnati* by expressly granting rate-setting power through the Hepburn Act of 1906[54] and Mann-Elkins Act of 1910.[55] Courts, meanwhile, began distinguishing legal questions (their province) from substantive technical analysis (the agency's domain).[56] Second, *procedural review*: In *ICC v. Louisville & Nashville Railroad*, a 1913 rate discrimination case, the Supreme Court held that while courts could not review an agency's substantive conclusions in an adjudication, they could review the agency's record of administrative proceedings to determine whether the agency had acted based on sufficient evidence and with sufficient procedural protections.[57] Taken with *Illinois Central*'s injunction that courts should not question the "wisdom" of agency actions[58] unless "so arbitrary and unreasonable as to render it invalid,"[59] these cases created procedural review to compensate for the courts' declining capacity for substantive review. Third, *information-forcing*: In *ICC v. Goodrich Transit Co.*, the Court upheld ICC orders requiring carriers to adopt uniform accounting and submit comprehensive reports, making the railroads more comprehensible to government.[60] Furthermore, under *Louisville & Nashville Railroad*, the ICC was required to provide information to regulated parties, holding hearings and making decisions based on written records to enable review.[61] Constitutional barriers to federal regulation also began to diminish, enabling a national administrative state.[62]

---

[51] 167 U.S. at 505–11. *See also* Thomas W. Merrill, *Rethinking Article I, Section I: From Nondelegation to Exclusive Delegation*, 104 COLUM. L. REV. 2097, 2111 (2004).

[52] 167 U.S. at 511–12.

[53] Professor Thomas W. Merrill specifically dates this appellate review model to judicial review of ICC orders around 1910. Thomas W. Merrill, *Article III, Agency Adjudication, and the Origins of the Appellate Review Model of Administrative Law*, 111 COLUM. L. REV. 939, 942 (2011).

[54] Ch. 3591, 34 Stat. 584 (1906). *See* McCraw, *supra* note 9, at 62.

[55] Ch. 309, 36 Stat. 539 (1910). *See* McCraw, *supra* note 9, at 62–63.

[56] 215 U.S. 452, 464–66 (1910). Unless the agency's action was so arbitrary or unreasonable that it was invalid. 215 U.S. at 471.

[57] 227 U.S. 88, 92–93 (1913).

[58] 215 U.S. at 470.

[59] *Id*.

This first settlement laid administrative law's foundations, but imperfectly. Statutory delegation could authorize expert government but not ensure that it served public rather than private interests. The substitution of procedure for substance meant that courts could ensure that agencies gave reasons for their decisions but not that those reasons were good. Transparency through information-forcing and statistical analysis still required expertise to interpret it. By 1920, the railroad settlement had created government capable of managing these new industrial technologies, but at the price of increasing distance from democratic control. By the end of the decade, the Great Depression demanded another round of governmental innovation.

### *B. Mass Society and Statistical Governance (1920–1946)*

Since the turn of the twentieth century, the United States had completed a transition from a fragmented, regional economy to a national one.[63] The economy was much more productive, complex, and interlinked,[64] making it sensitive to demand shocks. Some new instrumentalities of national economic regulation, including the Federal Trade Commission and Federal Reserve, emerged in this period.[65] However, especially after World War I, a succession of comparatively non-interventionist presidents mostly left the economy to run on its own, believing that the conditions for "a permanent plateau of prosperity" had been established.[66]

---

[60] 224 U.S. 194, 212–16 (1912).

[61] 227 U.S. at 104–06.

[62] *See* 230 U.S. 352, 381–82 (1913); 234 U.S. 342, 351–52 (1914).

[63] Chandler, Jr., *supra* note 35, at 5.

[64] DAVID A. HOUNSHELL, FROM THE AMERICAN SYSTEM TO MASS PRODUCTION, 1800–1932: THE DEVELOPMENT OF MANUFACTURING TECHNOLOGY IN THE UNITED STATES 289 (1984); Vaclav Smil, *Nitrogen Cycle and World Food Production*, 2 WORLD AGRICULTURE 9 (2011).

[65] Federal Reserve Act, ch. 6, 38 Stat. 251 (1913); Federal Trade Commission Act, ch. 311, 38 Stat. 717 (1914).

In 1929, their hopes were dashed and the Great Depression began.[67] The technologically-enabled mass society that had birthed the Roaring Twenties now magnified and spread the hurt of the Crash around the country. By 1933, more than 13 million workers were unemployed, around a quarter of the total American workforce.[68] Government lacked the tools to comprehend, much less correct, systemic collapse affecting millions simultaneously. The Great Depression was not merely an economic crisis but a new kind of capability crisis.

These new problems required mass government, new kinds of expertise, and novel statistical approaches. Macroeconomic stabilization required population-level measures rather than case-by-case relief, overproduction and underconsumption required sectoral intervention, and industry-wide labor disputes called for standardized fact-finding and program rules.[69] Legacy governance institutions were insufficient to handle these new challenges. Courts could only adjudicate backward-looking disputes between specific parties,[70] Congress lacked the technical expertise to calibrate monetary policy or agricultural production quotas to shifting market conditions, and the expert commissions were designed to handle specific sectors rather than coordinate recovery across interdependent industries.[71] State governments, which had traditionally handled relief for the poor, found their treasuries exhausted and their regulatory reach stopped at borders that economic distress freely crossed.

---

[66] ARTHUR M. SCHLESINGER, JR., THE CRISIS OF THE OLD ORDER, 1919-1933 155 (1957).

[67] *Id.* at 159.

[68] *Id.* at 440.

[69] *See* DAVID M. KENNEDY, FREEDOM FROM FEAR: THE AMERICAN PEOPLE IN DEPRESSION AND WAR, 1929–1945 131–60 (1999); JOHN MAYNARD KEYNES, THE GENERAL THEORY OF EMPLOYMENT, INTEREST AND MONEY 246–64 (1936).

[70] *See* Lon L. Fuller, *The Forms and Limits of Adjudication*, 92 HARV. L. REV. 353, 363–70, 394–404 (1978) (explaining how adjudication's party-presentation form limits its ability to address polycentric, system-wide problems and complex repercussions).

[71] *See* CONG. RSCH. SERV., ECONOMICS OF FEDERAL RESERVE INDEPENDENCE 9–10 (RL31056, Apr. 17, 2007) (explaining that because the legislative branch is not positioned to exercise day-to-day control over monetary policy, delegation structures tend to place operational control in expert agencies).

The modern administrative state coalesced and was vastly expanded in the New Deal to address these systemic challenges. More than 20 "alphabet agencies" were created in 1933 and 1934 alone, dealing with problems from rural electrification to agricultural price adjustment to the banking crisis.[72] The New Deal agencies were mostly staffed by specialists, experts in their fields who were able to grapple with the unique challenges each of them presented, but at a new level of scale and statistical instrumentation.

New technologies played a crucial role in enabling the New Deal institutions to operate effectively, as they had enabled the railroad commissions. The Social Security Administration used punch-card systems to register 26 million workers in its first four months and began distributing benefits to 222,000 elderly Americans by 1940.[73] The Agricultural Adjustment Administration's statistical surveys facilitated production controls that raised farm income by 50% between 1932 and 1935, rescuing rural America from collapse.[74] The Securities and Exchange Commission's mandatory disclosure regime and market surveillance restored enough confidence that new securities issues rose from near zero in 1933 to $3 billion by 1936.[75] These agencies operated at population scale, using statistical methods and technologies from punch cards to microfilm to transform abstract economic aggregates into concrete improvements in human welfare.

But, as had occurred in the railroad era, the scale and power of these new agencies led to concerns that they were growing unaccountable. Congress had delegated power to the agencies, often with little detail about how they should use it. The agencies began to be described as a "fourth branch" outside of the constitutional bounds of American government.[76] Substantive oversight by Congress and the courts became harder again. For example, when the Agricultural Adjustment Administration set crop quotas affecting millions of farmers, no individual could easily understand or challenge the sector-wide view and statistical models underlying production targets. Social Security's punch-card systems processed citizens as data points, transforming democratic subjects into statistical objects.

---

[72] *See* Bob Clark, *FDR, Archivist: The Shaping of the National Archives*, 38 PROLOGUE MAG. (Nat'l Archives, 2006).

[73] *See* Carolyn Puckett, *Administering Social Security: Challenges Yesterday and Today*, 70 SOC. SEC. BULL. 27, 29 (2010); *Historical Background and Development of Social Security*, SOC. SEC. ADMIN. (n.d.), https://www.ssa.gov/history/briefhistory3.html.

[74] *See* E.M. Brooks, *As We Recall: The History of Agricultural Estimates, 1933 to 1961*, U.S. DEP'T OF AGRIC. NAT'L AGRIC. STAT. SERV. (1977).

[75] SECURITIES AND EXCHANGE COMMISSION, THIRD ANNUAL REPORT (FISCAL YEAR ENDED JUNE 30, 1937) 12, (reporting "new securities proposed to be offered for cash" of approximately $2.84 billion during the fiscal year).

The initial accountability pushback occurred in the courts. In *Panama Refining*[77] and *Schechter Poultry*,[78] two cases from 1935, the Supreme Court struck down parts of the National Industrial Recovery Act on the basis that it contained unconstitutional delegations of legislative power and also exceeded congressional authority under the Commerce Clause. This "nondelegation" doctrine echoed earlier concerns about the delegation of legislative power in *ICC v. Cincinnati* but went further in its limitations on congressional delegation.[79] In *Carter v. Carter Coal Co.* and *United States v. Butler*, the Court limited Congress's basic ability to regulate federally under the Commerce Clause and Spending Power.[80]

Despite this pushback, the elected branches demanded the administrative state, leading to the threat of court packing and the "switch in time" of 1937.[81] That year, the Court upheld a state minimum wage in *West Coast Hotel*[82] and read the Commerce Clause expansively in *Jones & Laughlin*

---

[76] THE PRESIDENT'S COMM. ON ADMINISTRATIVE MANAGEMENT, ADMINISTRATIVE MANAGEMENT IN THE GOVERNMENT OF THE UNITED STATES 30 (1937); FTC v. Ruberoid Co., 343 U.S. 470, 487 (1952) (Jackson, J., dissenting) ("[A]dministrative agencies have become a veritable fourth branch of the Government").

[77] 293 U.S. 388 (1935).

[78] 295 U.S. 495 (1935).

[79] *See* 293 U.S. at 430 (establishing the nondelegation doctrine).

[80] 298 U.S. 238, 303–08 (1936); 297 U.S. 1, 68–69 (1936).

[81] *See* 2 BRUCE ACKERMAN, WE THE PEOPLE: TRANSFORMATIONS 279–344 (1998).

[82] 300 U.S. 379, 391–97 (1937).

*Steel*.[83] By 1944, *Yakus v. United* States blessed price controls on delegations that could not have survived the nondelegation test of a decade prior.[84]

But worries of government overreach and abuse persisted, especially once the emergencies of the Great Depression and World War II receded. The courts had retreated from substantive oversight, even as the administrative state pressed against constitutional constraints. Statistical governance made it difficult to ensure due process to the millions now subjected to regulation, and courts lacked capacity to review every breach given the rate and complexity of agency actions.[85] A new doctrinal framework was needed. The solution would be to further substitute procedural regularity for substantive accountability—if courts couldn't evaluate mass statistical decisions, they could at least ensure proper processes had been used.

The Administrative Procedure Act (APA)[86] was passed in 1946 to provide the foundation for this new settlement. This "super-statute"[87] structures administrative law to this day, applying to all agencies except to the extent that their authorizing statute supersedes it.[88] Its key provisions expand on the allocation, review, and information-forcing dimensions of the railroad era's administrative settlement. First, the APA provides the basic structure for the roles played by administrative agencies and by courts. Agency actions, now statutorily classified as either rulemakings or adjudications, are subject to a baseline expectation of judicial review with specific standards to govern it.[89] Second, each of these kinds of agency action must follow an enumerated set of formal and informal procedures, including decision-making on a record, which provide the basis for procedural review of agency action.[90] Third, the APA established procedural transparency requirements for agency actions that, when combined with caselaw holding that agencies had to make decisions actually based on the evidentiary record in front of them, operate as an information-forcing mechanism to bring administration further into the public and judicial view.[91]

---

[83] 301 U.S. 1, 37 (1937).

[84] 321 U.S. 414, 423–27 (1944).

[85] *See* Jerry L. Mashaw, *The Management Side of Due Process: Some Theoretical and Litigation Notes on the Assurance of Accuracy, Fairness, and Timeliness in the Adjudication of Social Welfare Claims*, 59 CORNELL L. REV. 772 (1974) (analyzing due process constraints under conditions of mass administration).

[86] Administrative Procedure Act, Pub. L. No. 79-404, 60 Stat. 237 (1946) (codified as amended at 5 U.S.C. §§ 551–559, 701–706).

[87] *See generally* William N. Eskridge Jr. & John Ferejohn, *The APA as a Super-Statute: Deep Compromise and Judicial Review of Notice-and-Comment Rulemaking*, 98 NOTRE DAME L. REV. 1893 (2023).

[88] 5 U.S.C. § 559.

[89] 5 U.S.C. §§ 701–06.

Taken together, these elements of the APA allowed the population-level governance necessary to handle the challenges raised by the Great Depression and a national economy and society while preserving accountability and guarding against the worst abuses of administrative power. Rather than restore individual-level accountability, impossible under statistical governance, the APA created procedural protections that aim to protect individual rights and seek to incentivize correct agency decisions through mandating process. The APA could not make mass administration comprehensible to the general public but gave it procedural legitimacy instead.[92] Statistical and expert governance were reconciled with the demands of legality. The APA settlement remains the foundation for the administrative state today.

However, the APA settlement again came with a cost of increased complexification and distancing from the public. Statistical governance became permanent, procedural complexity became mandatory, and democratic accountability became increasingly mediated through technical processes. Government grew significantly, in terms of the number of agencies, personnel, rulemakings and adjudications, and the expertise required to engage with them. In the decades to come, environmental and safety crises would again demand a technologically upgraded government, destabilizing the APA settlement and paving the way for another cycle of unsettlement.

---

[90] 5 U.S.C. §§ 553–58.

[91] *See* 5 U.S.C. § 552(a)(1)–(2) (publication/availability requirements); id. § 706 (record-based review standards); SEC v. Chenery Corp. (Chenery I), 318 U.S. 80, 87 (1943) (agency action must stand or fall on the grounds invoked by the agency in the record).

[92] *See* JERRY L. MASHAW, DUE PROCESS IN THE ADMINISTRATIVE STATE 1–31 (1985) (explaining proceduralized legality as the central currency of legitimacy in mass administration).

### *C. Computers and Complex Science (1946–1985)*

The New Deal had mostly focused on concrete, population-level harms like unemployment or labor abuses, but new and more diffuse problems would soon emerge. The new environmental regulation of the 1960s and 70s responded to problems like water, air, and soil pollution which were being worsened by synthetic pesticides and similar inventions.[93] These problems demanded more sophisticated scientific understanding, measurement, and prediction than earlier forms of administration could provide. Regulators had to identify issues in complex biological systems, anticipate secondary effects and tradeoffs, and design implementation frameworks that could operationalize scientific judgments at scale.[94]

Beginning in the early 1960s, Congress passed laws obliging agencies to resolve these scientific problems. The Clean Air Act,[95] Clean Water Act,[96] National Environmental Policy Act (NEPA),[97] Occupational Safety and Health Act (OSHA),[98] and others presented new kinds of questions. What concentration of lead endangers public health? What is a "significant" carcinogenic risk in the workplace? What probability of containment failure is acceptable at a nuclear facility? Congress could not legislate specific answers, both because it lacked substantive expertise and because if it tried to legislate specifically, new evidence or shifting technology might render the statutory answer obsolete in practice but mandated in law. These were problems legible only through instrumentation and simulation, yet shot through with normative questions that could not simply be handed to experts.[99]

---

[93] *See* Jack Lewis, *A Historical Perspective on Environmental Regulations*, EPA J. (Mar. 1988), https://www.epa.gov/archive/epa/aboutepa/looking-backward-historical-perspective-environmental-regulations.html.

[94] *See* Sheila Jasanoff, *Science, Politics, and the Renegotiation of Expertise at EPA*, 7 OSIRIS 194, 197–205 (1992).

[95] Clean Air Act § 109(b)(1), 42 U.S.C. § 7409(b)(1).

[96] Clean Water Act § 304(a)(1), 33 U.S.C. § 1314(a)(1).

[97] NEPA § 102(2)(C), 42 U.S.C. § 4332(2)(C).

[98] Occupational Safety and Health Act § 6(b)(5), 29 U.S.C. § 655(b)(5).

[99] *See* Alvin M. Weinberg, *Science and Trans-Science*, 10 MINERVA 209, 209–215 (1972) (arguing that many public problems are scientific, political, and also "trans-scientific" in that they deal with questions that scientific experiments cannot answer).

Agencies had to adopt new technologies to respond to these problems. Statistical models of complex systems including atmospheric transport, exposure-response curves, and probabilistic risk assessment were needed to estimate harms and regulatory impacts.[100] Building and analyzing large datasets required new techniques of measurement and the expansion of computing capacity. The Environmental Protection Agency implemented dispersion models and multi-pollutant inventories.[101] The Nuclear Regulatory Commission piloted probabilistic risk assessment at the scale of $10^{-6}$ chance of harm occurring.[102] The Securities and Exchange Commission began electronic market surveillance that could keep up with computerized trading.[103] These models created regulatory facts through computerized analysis. When EPA's atmospheric transport models predicted pollution concentrations, their outputs were legally binding despite resting on layers of contested modeling assumptions. Computing both increased and limited the scrutability of administrative action. Runs of models, parameter files, and validation studies could be filed, audited, and contested, but were still more difficult for non-specialists, including courts and legislatures, to understand than simpler earlier models.[104]

These capability gains again created accountability problems. Agencies operated in specialized domains where courts could not even fulfill their traditional function of using statutory interpretation to guarantee the legality of administrative action. Many terms in authorizing statutes, from the Clean Air Act to OSHA, were highly technical and diverged from ordinary usage.[105] The era's debates over statutory interpretation reflected anxiety on the part of judges as to whether they would be able to actually interpret the laws such that they could sufficiently constrain agency action.[106] In the *Chevron* deliberations in 1984, Justice John Paul Stevens reportedly admitted that "when I get so confused, I go with the agency," an admission that the complexity of modern administration had exceeded the capacity of the courts to oversee it.[107]

---

[100] *See, e.g.,* 40 C.F.R. pt. 51, app. W § 1.0(b)–(d) (explaining that air monitoring data alone are often insufficient and impacts of new sources "can only be determined through modeling").

[101] *Id.*

[102] *See* Weinberg, *supra* note 99, at 210 (discussing such risk modeling).

[103] SECURITIES AND EXCHANGE COMMISSION, ANNUAL REPORT 1981 11–12 (1981).

[104] Jasanoff, *supra* note 94, at 203, 209–11 (describing dissonance between public demands for "untainted knowledge" and the reality that policy-relevant science assimilates to politics, and highlighting contestation over models and underlying assumptions).

[105] *See* Ethyl Corp. v. EPA, 541 F.2d 1, 36–38 (D.C. Cir. 1976) (emphasizing that in "technical" cases a reviewing court must work to "understand enough about the problem confronting the agency" to perform its appellate function, while insisting review ultimately asks for a "rational basis" rather than judicial substitution of judgment).

Procedural protections could substitute for substantive review only in part. The gap between agency action and review often stretched to years, meaning that unless courts enjoined agency interventions (often burdening the parties the government was supposed to protect), remedies were far away.[108] Where courts relied on increased proceduralization to protect disempowered groups, resource asymmetries emerged as sophisticated parties marshaled resources and experts to navigate procedures and influence agency action that under-resourced groups could not.[109] This "capture" dynamic and the flowering of public choice theory would become key influences in attempts to rein in the administrative state.[110] Another resource asymmetry emerged between agencies, with large staffs and high technology, and courts, which still ran as small organizations and often on paper.[111] Thus, once capture had occurred or the agency otherwise grew misaligned, it was difficult for the judiciary to identify and correct it.

Administrative law again sought a new settlement of capability and accountability, relying again on the three tools of allocation of authority, review, and information-forcing. This new settlement built on the previous ones, but a few key innovations stand out. First, courts fine-tuned review of agency actions. The administrative record became the basis for review, and the "hard look" doctrine established in the 1971 case *Citizens to Preserve Overton Park v. Volpe* required that courts consider the whole record to determine whether the agency had acted based on "all relevant factors" without a "clear error in judgment."[112] *Motor Vehicle Manufacturers Association v. State Farm* held that courts would use the record to assess whether agencies had considered important aspects of the problem, addressed reasonable alternatives and counterarguments, and gave "adequate reasons" for changes.[113] The effect was to proceduralize rationality, expanding the role of the courts into evaluating how agency reasoning occurred rather than if it was correct.

[106] *See* Indus. Union Dep't, AFL-CIO v. Am. Petroleum Inst. (The Benzene Case), 448 U.S. 607, 652–58 (1980).

[107] Mark Sherman, *High Court Climate Case Looks at EPA's Power*, HUFF POST (Feb. 23, 2014, 3:59 PM), https://perma.cc/SQ6K-5ANA.

[108] *See* Nicholas Bagley, *The Procedure Fetish*, 118 MICH. L. REV. 345, 350, 360–63 (2019) (explaining how procedural demands and litigation risk can delay or derail protective regulation and shift policymaking forms).

[109] *See* Marc Galanter, *Why the "Haves" Come Out Ahead: Speculations on the Limits of Legal Change*, 9 LAW & SOC'Y REV. 95, 97–104 (1974).

[110] *See* George J. Stigler, *The Theory of Economic Regulation*, 2 BELL J. ECON. & MGMT. SCI. 3, 3–5 (1971); Daniel A. Farber & Philip P. Frickey, *The Jurisprudence of Public Choice*, 65 TEX. L. REV. 873, 880–82 (1987).

[111] JOHN G. ROBERTS, JR., 2023 YEAR-END REPORT ON THE FEDERAL JUDICIARY 2–5 (charting the history of technology in the judiciary).

Second, deference doctrine crystallized as a form of allocating authority and oversight to preserve both capabilities and accountability. *Chevron U.S.A. Inc. v. NRDC*, holding that if Congress had not directly spoken to an issue of interpretation, courts should defer to agencies' reasonable construction of ambiguous statutes, became the canonical statement of interpretive deference. When the Clean Air Act used "stationary source," courts faced a term whose meaning emerged from engineering practice, atmospheric chemistry, and economic modeling, domains where legal training provided no insight.[114] In 1983, in *Baltimore Gas & Electric Co. v. NRDC*, the Court upheld the Nuclear Regulatory Commission's assumption that permanent nuclear waste storage would have no significant environmental impact because courts should be deferential to agencies under conditions of uncertainty such as estimating the progress of future technology.[115] Combined with *Chevron*, these cases meant that courts would defer substantive questions, including interpretation of ambiguous statutes, to agencies, and would police procedure using the tools of hard look review. Courts had retreated again from evaluations of substance, again seeking in procedure a partial substitute that could prevent lawless administration.

Finally, a variety of legislative, judicial, and executive actions worked to make the informational underpinnings of the administrative state legible to review and oversight through information-forcing. The Freedom of Information Act of 1966[116] (FOIA) and the Government in the Sunshine Act[117] enabled public review of agency files across the administrative state. NEPA[118] and the D.C. Circuit case *Calvert Cliffs' Coordinating Committee v. AEC* institutionalized environmental impact statements and alternatives analysis, forcing agencies to publicize and defend the scientific bases of their actions.[119] The Paperwork Reduction Act of 1980 established the Office of Information and Regulatory Affairs (OIRA) within the Office of Management and Budget to review agency actions.[120] Executive Order 12,291 centralized regulatory review and required cost-benefit analysis.[121] These information-forcing mechanisms did not rely upon courts or Congress to develop the technical expertise to govern the substance of agency actions, but rather empowered interested parties to understand and, if necessary, object to the workings of government.

---

[112] 401 U.S. 402, 416, 420 (1971).

[113] 463 U.S. 29, 42–44 (1983).

[114] 467 U.S. at 840–41 (describing the "bubble concept" dispute over the meaning of "stationary source"); *see* Richard J. Pierce, Jr., *What Actually Happened in Chevron*, YALE J. REG. NOTICE & COMMENT (Jan. 23, 2018), https://www.yalejreg.com/nc/what-actually-happened-in-chevron-by-richard-j-pierce-jr/.

[115] 462 U.S. 87, 103 (1983).

The Supreme Court did seek to limit this mounting proceduralization in *Vermont Yankee Nuclear Power Corp. v. NRDC*[122] and *Heckler v. Chaney*.[123] These limitations proved insufficient.[124] Procedure is an imperfect substitute for substance, and while the courts were doing what they could to extend the doctrinal settlement of allocation, review, and information to deal with the challenges of complex science, these moves came at the cost of procedural intensification and slower government, which hampers administration today.[125]

---

[116] Freedom of Information Act, Pub. L. No. 89-487, § 1, 80 Stat. 250, 250–52 (1966).

[117] Government in the Sunshine Act, Pub. L. No. 94-409, § 3(a), 90 Stat. 1241, 1241–49 (1976) (codified at 5 U.S.C. § 552b).

[118] National Environmental Policy Act of 1969, Pub. L. No. 91-190, § 102(2)(C), 83 Stat. 852, 852–53 (1970) (codified as amended at 42 U.S.C. § 4332(2)(C)).

[119] 449 F.2d 1109, 1112–14 (D.C. Cir. 1971).

[120] 44 U.S.C. § 3503(a)–(b) (creating OIRA within OMB); id. § 3504(a) (to "oversee the review and approval of information collection requests," federal statistical activities, records management, and use of information technology).

[121] Exec. Order No. 12,291, §§ 2(b)–(d), 3(a), 3(c)–(e), 46 Fed. Reg. 13,193, 13,193–95 (Feb. 19, 1981).

[122] 435 U.S. 519, 524–25 (1978).

[123] 470 U.S. 821, 831–35 (1985).

[124] Bagley, *supra* note 108, at 350–55, 360–64; Thomas O. McGarity, *Some Thoughts on "Deossifying" the Rulemaking Process*, 41 DUKE L.J. 1385, 1387–93 (1992).

[125] McGarity, *supra* note 124, at 1385–87; Richard J. Pierce, Jr., *Rulemaking Ossification Is Real: A Response to "Testing the Ossification Thesis"*, 80 GEO. WASH. L. REV. 1493, 1493–96 (2012).

By 1985, the administrative state had accumulated three layers of complexity, each persisting alongside its successors: sectoral expert regulation in the commission tradition; mass statistical administration for social programs; and model-driven risk governance for environmental and technological hazards. A single major EPA regulatory action could now require deep technical records and expert judgment, population-scale statistical inference and risk assessment, and computerized modeling. It also came with procedural requirements from all three eras that had likewise layered over time: APA notice-and-comment, record-based "hard look" judicial review, and (for executive agencies) centralized OMB review and cost–benefit analysis under E.O. 12,291. The sedimentary layers had compressed into governmental geology that no single actor could fully comprehend, and the balance between capability and accountability grew unsteady.

## II. A CRISIS OF ACCOUNTABILITY

For a hundred years, American administrative law followed a clear pattern: new technologies created capability crises, government complexified in response, accountability gaps emerged, and doctrinal settlements restored balance. Procedural oversight through administrative law created sufficient accountability at each step. Today, that pattern has broken. The administrative state is being dismantled without a precipitating technological crisis—by the Supreme Court through doctrine and by an executive branch pursuing its own program of simplification through mass workforce reductions, the elimination of entire agencies, and the reclassification of career officials for at-will removal.[126] New technologies like the internet and narrow AI have emerged since the *Chevron* settlement but did not fundamentally transform government in such a way as to create a capability crisis. A backlash has emerged nonetheless, this novel accountability crisis threatening to push administration back past the New Deal.[127]

This discontinuity demands explanation. This Part argues that administrative law has collapsed not from new technological disruption but from accumulated procedural weight, each settlement adding procedural requirements without removing previous ones. The result is a double deficit. The federal workforce is smaller *per capita* today than it was in the 1960s even as demands have expanded.[128] Far from alleging that government is doing too much, complaints that government cannot do anything have become a key part of contemporary political discourse.[129] Government has become both less capable and less accountable, throttled by procedure while simultaneously captured by those sophisticated enough to navigate it.

[126] *See supra* note 7.

[127] *See* Gillian E. Metzger, *Foreword: 1930s Redux: The Administrative State Under Siege*, 131 HARV. L. REV. 1, 1–6 (2017) (describing the rising tide of anti-administrative sentiment in the Supreme Court).

Faced with this crisis, the Supreme Court has chosen a blunt remedy: shrinking administration back to comprehensible size. This Part examines the causal chain that produced this crisis: first, how accumulated procedural complexity transformed administration into a bargaining arena that advantages repeat players and resists reform; second, how this dynamic explains the failure of internet-era technology to transform government; and finally, how the resulting legitimacy crisis triggered the current judicial retrenchment.

### *A. The Sedimentary State*

The procedural accumulation described in Part I both made government increasingly complicated and changed the character of administration. The traditional model of administrative action was of neutral, scientific experts simply carrying out the will of Congress through their specialized capabilities.[130] However, this "transmission belt" model relied upon a separation between politics and expertise that did not accurately describe how administrators exercised their discretion, undermining the legitimacy of scientific administration.[131] By the early 1970s, a new justification for administration arose, arguing that agency actions operated as pluralist forums in which organized interests could make their voices heard through legally structured procedures.[132] Groups with strong preferences could invest more in participation, and procedures became rules of these "markets" rather than guarantors of the fairness of administrative action.

[128] *The evolution of government employment*, FRED BLOG (Dec. 30, 2024), https://fredblog.stlouisfed.org/2024/12/the-evolution-of-government-employment/; U.S. GOV'T ACCOUNTABILITY OFF., INTERNAL REVENUE SERVICE: ABSORBING BUDGET CUTS HAS RESULTED IN SIGNIFICANT STAFFING DECLINES AND UNEVEN PERFORMANCE (GAO-14-534R) (Apr. 2014).

[129] *See generally* EZRA KLEIN & DEREK THOMPSON, ABUNDANCE (2025).

[130] Richard B. Stewart, *The Reformation of American Administrative Law*, 88 HARV. L. REV. 1667, 1671–75 (1975).

[131] *Id.* at 1776–81.

In this model, regulatory outputs are legitimated by structured participation and judicially policed reason-giving rather than by claims of technocratic neutrality. Because Congress could no longer translate the will of the people into comprehensive guides for agency action,[133] the popular will would be expressed through pluralist bargaining. Judicially-governed process provided the framework. Notice-and-comment opens the door to informed outsiders, an expanded administrative record makes choices contestable, and "hard look" review enforces responsiveness to salient criticisms. Courts would defer under *Chevron* because they could not credibly resolve scientific questions themselves, and substituting their judgment would undermine government's legitimacy on another dimension. Administrative agencies must decide questions that, while they involve science, are fundamentally normative and subject to the discretion of administrative officials, like what counts as "fair" or "unhealthy."[134] Political participation through bargaining where different interests are represented reduces the extent to which these questions are insulated from public input.

However, the agency forums are not truly representative. Repeat players have the resources and familiarity with the regulatory process to exert disproportionate influence over agency decisions.[135] McCubbins, Noll, and Weingast reframe the procedural moves that supported this shift as deck-stacking, institutional design that embeds legislative compromises and distributes leverage across favored constituencies.[136] McCubbins and Schwartz argue that Congress deliberately relies on fire-alarm oversight, deputizing organized groups to monitor agencies through those procedures instead of conducting continuous "police-patrol" oversight itself.[137] In domains with a high degree of complexity of agency action, this kind of expert interest group oversight can compensate for substantive congressional incapacity. Yet the groups best positioned to pull the fire alarm are those with resources to monitor agencies continuously, not the diffuse public that agencies are meant to serve.

[132] *Id.* at 1711–13.

[133] *Id.* at 1671–75.

[134] Jasanoff, *supra* note 94, at 203–06.

[135] Stewart, *supra* note 130, at 1713–15.

[136] McCubbins, Noll, & Weingast, *supra* note **Error! Bookmark not defined.**, at 249–53.

Because rules became tools for interest groups, procedures no longer simply maintain rule of law. Instead, they shape bargaining domains, and shifting them advantages or diminishes specific parties. Notice-and-comment rulemaking is formally an information-gathering device but practically operates as an agenda-shaping and litigation-positioning stage.[138] Repeat players file submissions that target the fulcrum assumptions of an agency's analysis and lay the groundwork for *State Farm* challenges if the final rule does not squarely address reasonable alternatives. Agencies respond strategically, pre-negotiating with likely litigants, over-documenting "consideration of relevant factors," and drafting response-to-comment matrices with an eye toward remand risk rather than clarity.[139] Each procedural requirement, defensible for external reasons in isolation, becomes another surface on which sophisticated parties can apply pressure. The cumulative effect is ossification, the increasing structural paralysis of government,[140] and increased inscrutability.

This dynamic creates a two-sided legitimacy problem. As veto points proliferate and administrative reasons are crafted to withstand litigation rather than to be genuinely transparent to the public, the outputs of administrative action look like lowest common denominator compromises rather than what is genuinely in the public interest. The distributional fights that often are the main focus of conflict in these areas do not seem well-governed by the judicial procedural review that is intended to give administration legitimacy. External legitimacy, in the form of presidential administration, can provide needed energy and a democratic justification for the actions of the administrative state.[141] However, even with OIRA and similar forms of executive review, much of agency action remains outside of presidential control because of its size and complexity.[142]

[137] Mathew D. McCubbins & Thomas Schwartz, *Congressional Oversight Overlooked: Police Patrols versus Fire Alarms*, 28 AM. J. POL. SCI. 165, 166 (1984).

[138] *See* McCubbins, Noll, & Weingast, *supra* note **Error! Bookmark not defined.**, at 258–60.

[139] *See* Bagley, *supra* note 108, at 351–52, 360–61.

[140] *See* McGarity, *supra* note 124, at 1386–90.

More fundamentally, pluralist bargaining is not representative. Influence shifts from diffuse citizens to repeat players who can credibly threaten delay, remand, or appropriations retaliation.[143] The construction company that will litigate an EPA rule has a seat at the table that the asthmatic children who will breathe the air do not. Pluralist bargaining allows public input on key normative questions that agencies must decide, but when only certain players can provide input effectively, its utility is undermined. Administrative law, which is intended to be the fundamental democratic check on an increasingly powerful, complicated, and expert government becomes a kind of politics by other means which does not even have the virtue of complete democratic representation.

Attempts to reform the system disadvantage particular interest groups in bargaining, leading to resistance, and so procedure piles up while capacity and accountability erode. This sets the stage for Part II.B's puzzle: why, unlike every prior era, did the transformative new technologies of the internet and narrow AI fail to transform government?

## *B. Technological Stagnation*

The technologies developed since 1985 have transformed modern life, but mostly not modern government. In 1990, less than one percent of Americans used the internet.[144] By 2023, that number had reached 93%,[145] as the World Wide Web, email, the smartphone, and social media pushed life online.

---

[141] *See* Kagan, *supra* note **Error! Bookmark not defined.**, at 2341.

[142] *See* Bednar, *supra* note **Error! Bookmark not defined.**, at 830–31.

[143] Galanter, *supra* note 109, at 97–104.

[144] World Bank, *Internet users for the United States* (FRED series ITNETUSERP2USA) (reporting 1990: 0.7850 internet users per 100 people).

[145] *Id.* (reporting 2023: 93.1444 internet users per 100 people).

Yet agencies underwent only superficial digitalization. The E-Government Act of 2002 mandated online services and[146] Regulations.gov digitized comments.[147] Electronic filing systems, including in the U.S. Patent and Trademark Office,[148] the Securities and Exchange Commission's EDGAR,[149] and the federal courts' PACER,[150] moved much information access online. But these incremental improvements did not constitute the kind of institutional transformation that had previously driven administrative evolution but rather a shift in how to access existing tools.[151] Even on its own terms, government uptake of technology has been relatively ineffective. As of 2025, the Government Accountability Office continues to list federal IT acquisition and management as a "high-risk" area, finding that investments "too frequently fail or incur cost overruns and schedule slippages while contributing little to mission-related outcomes," notwithstanding annual federal IT spending exceeding $100 billion.[152] Crises from the failure of HealthCare.gov[153] to the scramble to find COBOL coders to fix unemployment systems that crashed during the COVID-19 pandemic demonstrate government's underlying incapacity.[154]

---

[146] E-Government Act of 2002, H.R. 2458, 107th Cong. (as enrolled).

[147] *See Presidential Initiatives: E-Rulemaking*, GEORGE W. BUSH WHITE HOUSE (archives.gov) (n.d.), https://georgewbush-whitehouse.archives.gov/omb/egov/c-3-1-er.html.

[148] Zandra Smith & Lisa Tran, *Filing for a Patent Using EFS Web*, U.S. PATENT AND TRADEMARK OFFICE (2017), https://www.uspto.gov/sites/default/files/documents/Website%20PDF%20-%20Invention%20Con%202017%20Filing%20for%20a%20Patent%20-%20OPLA.pdf.

[149] Mauri L. Osheroff, Mark W. Green, & Ruth Armfield Sanders, *Electronic Filing and the EDGAR System: A Regulatory Overview*, U.S. SECURITIES AND EXCHANGE COMMISSION (Oct. 3, 2006), https://www.sec.gov/info/edgar/regoverview.htm.

[150] *About Us*, PUBLIC ACCESS TO COURT ELECTRONIC RECORDS (PACER) (n.d.), https://pacer.uscourts.gov/about-us.

[151] As Jane E. Fountain predicted might happen in her pathbreaking book, "reorganizing and restructuring the institutional arrangements in which [government to citizen] transactions are embedded" proved to be much harder than simply taking up new technology. JANE E. FOUNTAIN, BUILDING THE VIRTUAL STATE: INFORMATION TECHNOLOGY AND INSTITUTIONAL CHANGE 6–7 (2001).

[152] U.S. Gov't Accountability Off., *High-Risk Series: Critical Actions Needed to Urgently Address IT Acquisition and Management Challenges* 2–4 (GAO-25-107852, 2025).

[153] JENNIFER PAHLKA, RECODING AMERICA: WHY GOVERNMENT IS FAILING IN THE DIGITAL AGE AND HOW WE CAN DO BETTER 118–134 (2023).

[154] Michael A. Navarrete, *COBOLing Together UI Benefits*, at 1–2 (Brookings Working Paper 98, 2024).

Instead of reducing the need to rely on procedures developed for a prior technological era, and so making government more scrutable, the internet left government much as it was—often mediated by private actors providing interfaces optimized for their own purposes.

Why have these technologies not transformed government? Institutional factors, like civil service rules, procurement processes, risk aversion, and inadequate appropriations, present obstacles. Fundamentally, however, proceduralization and the rise of bargaining through the sedimentary state discussed in Part II.A provide a deeper explanation for why transformation has not worked well.

Procedure disincentivizes innovation within government. An agency redesigning operations would need to alter workflows formalized in regulations, change processes organized interests have optimized around, and modify record-keeping that forms the basis for judicial review. Each change would create winners and losers among the constituencies that have learned to navigate existing procedures, many of whom would bring legal challenges arguing that the new process violates statutory requirements or denies procedural rights. Each change would require approval through the same ossified channels that Part II.A described: interagency review, OIRA clearance, notice-and-comment on any rule changes, and anticipation of litigation from those disadvantaged by reform. As such, procedure has not just prevented government from effectively carrying out its duties in many domains but also from upgrading its ability to do so, incapacity reaching to the roots.

TurboTax and the problem of tax reform illustrate the dynamic. Efforts to build simpler tax filing systems, a useful technology that would make government more accessible to ordinary citizens, have been met with lobbying efforts by established players whose business models depend on tax complexity.[155] The ability of private companies to uptake technology and fulfill quasi-governmental functions has created a constituency that is opposed to more effective government which might displace them. The same pattern appears across domains. As such, the procedural sediment described in Part II.A is not merely inert but actually defended by those who have adapted to it.

[155] Lawrence Zelenak, *Complex Tax Legislation in the TurboTax Era*, 1 COLUM. J. TAX L. 91, 92–95, 97–100 (2010).

Internet-era technological participation has amplified the surface dynamics without changing the underlying structure. Mass comments increase volume and noise, but agencies rationally prioritize the credible-litigant subset of those who have commented. A proposed rule might generate a million form comments from an advocacy campaign, but the agency's lawyers will focus on the detailed technical submissions from regulated industries that preview the arguments likely to appear in a petition for review.[156] Online dockets make agency proceedings more transparent, but transparency without capacity for engagement merely lets citizens watch a game they cannot play.

Agencies have failed to undergo the kind of fundamental process redesign that digital technology enables elsewhere. While some scholars have tried to develop visions for such transformation through the internet or narrow AI systems deployed in administrative domains, their visions of genuine transformation remain unrealized.[157] That may at last be changing. Since 2025, the executive branch has directed agencies to accelerate AI adoption. OMB's government-wide guidance requires Chief AI Officers, public use-case inventories, and minimum risk-management practices for "high-impact" systems. By spring 2026, agencies had reported more than 3,600 AI use cases, several hundred designated high-impact.[158] But reports of procurement failures and ineffective integration suggest that fundamental transformation here might not be forthcoming. And the use of AI in administration presents serious risks both to legitimacy, because of the black box nature of many systems,[159] and of abuse, such as biased algorithms.[160] However, these problems may be more correctable in AI systems than they are in humans,[161] and government will have to adopt new technology in order to keep up with the rapid advances made in the private sector.

---

[156] *See* Nina A. Mendelson, *Foreword: Rulemaking, Democracy, and Torrents of E-Mail*, 79 GEO. WASH. L. REV. 1343, 1346–50, 1359–76 (2011).

[157] *See, e.g.,* Cary Coglianese & David Lehr, Regu*lating by Robot: Administrative Decision Making in the Machine-Learning Era*, 105 GEO. L.J. 1147, 1167–68, 1174–75, 1180–81 (2017) (describing "adjudicating by algorithm" and "rulemaking by robot," and explaining how ML could be nested within larger models to support "automated rulemaking"); Anthony J. Casey & Anthony Niblett, *The Death of Rules and Standards*, 92 IND. L.J. 1401, 1403–05, 1410–14 (2017) (introducing "micro-directives," machine translation of legislative goals into "simple commands," and the claim that micro-directives can "update[] as quickly as conditions change"); David Freeman Engstrom & Daniel E. Ho, *Algorithmic Accountability in the Administrative State*, 37 YALE J. ON REGUL. 800, 800–02, 804–06, 849–54 (2020) (proposing "prospective benchmarking" to "learn about, validate, and correct errors" in algorithmic decisionmaking).

As agencies fall further behind the technological frontier, the gap between governmental and private-sector capabilities widens. Regulated entities operate with tools that regulators cannot match, and the rise of advanced AI threatens to make this problem even worse. The resource asymmetry between agencies and sophisticated parties that Part II.A identified as a feature of pluralist bargaining grows more severe when that asymmetry includes technological capacity. Agencies that cannot understand how regulated entities operate cannot effectively oversee them, and the procedural substitutes for substantive oversight that Part I traced become even more inadequate.

This account helps resolve the apparent puzzle of technological stagnation in government. Unlike prior eras, no capability crisis preceded the Supreme Court's current doctrinal unsettlement because government never gained the capabilities that new technologies could have provided. Procedures pile up, but capability does not accumulate alongside them. Government is left with the worst of both worlds: the complexity costs of three eras of procedural accretion without the technological benefits that might justify those costs. This paralysis helps explain why no new settlement or other significant transformation has emerged in administrative law despite four decades passing since *Chevron*. But the absence of a capability surge has not meant the absence of crisis. Instead, a different kind of crisis emerged, one driven not by governmental overreach but by governmental dysfunction. Part II.C examines how the Supreme Court has responded to this legitimacy vacuum.

---

[158] Office of Mgmt. & Budget, Exec. Office of the President, *2025 Federal Agency Artificial Intelligence Use Case Inventory* (updated Apr. 13, 2026) (reporting 3,611 individually reported AI use cases, including 445 high-impact use cases).

[159] *See* Deirdre K. Mulligan & Kenneth A. Bamberger, *Procurement as Policy: Administrative Process for Machine Learning*, 34 BERKELEY TECH. L.J. 773, 816–17 (2019) (explaining that ML can displace discernible reasons with models whose internal "reasoning" may be difficult or impossible to recover).

[160] *See* Julia Angwin et al., *Machine Bias*, PROPUBLICA (May 23, 2016), https://www.propublica.org/article/machine-bias-risk-assessments-in-criminal-sentencing; Joy Buolamwini & Timnit Gebru, *Gender Shades: Intersectional Accuracy Disparities in Commercial Gender Classification*, *in* PROC. FAT* 77, 80–83, 86–88 (2018).

[161] Jon Kleinberg et al., *Human Decisions and Machine Predictions*, 133 Q. J. ECON. 237, 251–57 (2018).

### *C. Dismantling for Administrative Simplification*

Parts II.A and II.B described a government that is often ossified, controlled by sophisticated repeat players, and unable to upgrade itself to handle the new problems society confronts. This is a crisis, but not the kind that prior cycles of doctrinal disruption and resettlement addressed. That traditional pattern involved capability surging ahead of accountability, as when the mass administration of the New Deal outpaced the ability of courts and legislatures to substantively review decisions. Here, both have eroded together, and over-capability has not been the driver of worries about governmental abuse.

Nevertheless, the United States is again undergoing a serious crisis in administrative law aimed at creating a greater degree of accountability in administration. The Supreme Court, in a series of recent decisions, has struck at the foundations of administrative agencies, creating what Professor Gillian Metzger calls "1930s Redux" as elements of administrative law settled since the New Deal come under attack.[162] The Court is not merely adjusting doctrine at the margins, as it does during the periods of settlement. Instead, it is dismantling parts of the fundamental architecture that resulted from the *Chevron* settlement, making significant changes to the same three tools of allocation, review, and information-forcing that enabled administrative law to mediate capability and accountability across a century of technological change. Nor is the Court acting alone. The Executive has pursued the political twin of the doctrinal project: reductions in force across the civil service, the shuttering of agencies, reclassification of career positions for at-will removal, and centralized control over spending and personnel.[163] These campaigns converge on the same end, an administrative state small and subordinate enough for a single principal to comprehend and control. As discussed below, the Court's removal jurisprudence has now constitutionalized the executive's half of the project.

[162] Metzger, *supra* note 127, at 1–3.

Increased partisanship[164] and a growing sense among the Justices that administrative agencies are captured by Democrats[165] may play a role, though the Court has had Republican-appointed majority since 1970 and *Chevron* itself was championed by Antonin Scalia.[166] But the framework developed in this Article suggests a structural explanation beyond partisanship: the accumulated complexity described in Parts II.A and II.B has made administration genuinely incomprehensible to generalist judges, and the Court is responding by trying to shrink government back to a size it can understand and therefore oversee. From this perspective, the Court can be understood to be seeking to restore accountability and democratic legitimacy to an administrative state that has been distanced from the popular will by complexity and pluralist bargaining.

This reading finds support in the Justices' recent decisions. Requiring explicit delegation from Congress on major questions, centralizing executive power in the President, and increasing the role of courts in overseeing administrative actions can each be seen as ways that the Court is seeking to make agencies answer to democratic authority, even at the cost of the output legitimacy that comes from effective, capable, and independent government. These moves also track the three doctrinal tools of settlement that Part I identified. And, in the Court's most recent term, they have reached past the tools to the institutional premise that those tools were built to discipline: insulated expertise itself.

---

[163] *See supra* note 7.

[164] Richard Re, *Foreword: To a Conservative Warren Court*, 139 HARV. L. REV. 1, 19 (2025); Metzger, *supra* note 127, at 15.

[165] Steven Teles, *It's time for liberals to confront their own anti-majoritarianism*, NISKANEN CENTER (Mar. 21, 2025), https://www.niskanencenter.org/its-time-for-liberals-to-confront-their-own-anti-majoritarianism/.

[166] *See* Antonin Scalia, *Judicial Deference to Administrative Interpretations of Law*, 1989 DUKE L.J. 511, 511–12 (1989).

First, the end of *Chevron* deference with the decision in *Loper Bright* transformed the allocation pillar of the prior administrative settlement.[167] Deference regimes like *Chevron* and *Auer*[168] had put much of legal interpretation into the hands of agencies who were deemed to be better equipped to understand how to resolve ambiguity in technical matters. In *Loper Bright*, the Court explicitly overruled *Chevron*, holding that "[c]ourts must exercise their independent judgment in deciding whether an agency has acted within its statutory authority" even where a statute is ambiguous.[169] Some degree of deference to agencies does persist through the background doctrine of *Skidmore* deference.[170] But overall the recent shift toward the use of the traditional tools of statutory interpretation by the judiciary to address these ambiguous problems, as *Loper Bright* and *Kisor v. Wilkie*[171] require, represents a more confident judicial posture towards technical interpretation and a shift in authority away from agencies toward the courts. Courts have reclaimed authority once ceded for want of understanding, and requiring that agencies act under judicial interpretations pushes administration back toward what courts can comprehend.[172]

Second, recent doctrinal shifts have broadened the set of areas in which courts can review agency authority. The expanding major questions doctrine demonstrates a new front in which judicial review of agency authority has opened up. Since the New Deal settlement, which ended the use of the nondelegation doctrine to police what Congress can and cannot delegate to the administrative state, delegations were doctrinally permissible if some "intelligible principle" guides the delegation.[173] Practically, however, no delegation has been struck down on the basis of this test since the New Deal, rendering it a dead letter. However, the major questions doctrine, which blocks delegations of "major question[s]" of "economic or political significance" to agencies unless there is clear congressional authorization for them to handle such issues, has become an increasingly active tool of limiting agency authority.[174] In *West Virginia v. EPA*, the Court deployed the major questions doctrine to invalidate EPA's Clean Power Plan, holding that the agency's interpretation of the Clean Air Act to allow generation-shifting as an emissions reduction strategy required clearer congressional authorization given its economic significance.[175] The effect is to require Congress to speak clearly on precisely the technical and contested questions that Congress delegated to agencies, attempting to force democratic deliberation over these questions even if Congress may struggle to effectively engage in it.

[167] 603 U.S. at 391–94, 412–13.

[168] 519 U.S. 452 (1997). *See* Thomas E. Nielsen & Krista A. Stapleford, *What Loper Bright Might Portend for Auer Deference*, HARV. L. REV. BLOG (Jul. 5, 2024), https://harvardlawreview.org/blog/2024/07/what-loper-bright-might-portend-for-auer-deference/.

[169] 603 U.S. at 394, 401–04, 412–13.

[170] *Id.* at 394, 403 (describing *Skidmore* respect and quoting "body of experience and informed judgment"). Skidmore v. Swift & Co., 323 U.S. 134, 140 (1944).

[171] 588 U.S. 558, 574–76 (2019).

[172] Whether this shift will allow for effective government is unclear. Forcing agencies to follow the interpretations of courts that do not understand the domains within which they operate could simply undermine their capacity to regulate effectively. *See* Vermeule, *supra* note 11, at 625–26.

Third, the Court has begun to restrict agency adjudication, narrowing the domain in which agencies can act as both prosecutor and judge and forcing more adjudications into the judicial branch. In *SEC v. Jarkesy*, the Court recognized a jury-trial right for SEC civil-penalty cases under the Seventh Amendment, curtailing in-house adjudication in that category.[176] This decision strikes at the adjudicatory capacity that has been central to agency enforcement since the New Deal; administrative courts with their own specialized judges and sets of procedures made adjudications more efficient while providing fairly strong procedural protections to defendants.[177] Agencies across domains may face similar challenges to their adjudicatory authority, potentially requiring them to litigate enforcement actions in courts already struggling with docket backlogs, creating more procedural protections for affected parties but potentially adding delay to an already sclerotic system.[178]

---

[173] J.W. Hampton, Jr., & Co. v. United States, 276 U.S. 394, 409 (1928) (articulating "intelligible principle"); Whitman v. Am. Trucking Ass'ns, 531 U.S. 457, 472–76 (2001).

[174] 597 U.S. at 721–24.

[175] 597 U.S. 697 (2022).

[176] 603 U.S. at 120–41.

[177] *See* CONG. RSCH. SERV., INFORMAL ADMINISTRATIVE ADJUDICATION: AN OVERVIEW (R46930 Oct. 1, 2021).

[178] The year before *Jarkesy*, in *Axon Enterprise, Inc. v. FTC*, the Court held that litigants in administrative proceedings could directly seek review of collateral constitutional claims about agency actions in the courts rather than having to first go through the agency appeals process. Axon Enters., Inc. v. FTC, 598 U.S. 175, 180 (2023).

Fourth, the Court has undone a keystone of independent administration. In *Trump v. Slaughter*, the Court overruled *Humphrey's Executor*,[179] holding that the FTC's for-cause removal protections violate the separation of powers.[180] "If anything more is left of *Humphrey's*," the Chief Justice wrote, "we overrule it."[181] Only the Federal Reserve, preserved "to the extent that it follows in the distinct historical tradition of the First and Second Banks of the United States," escapes.[182]

On the account developed here, *Slaughter* is the purest expression of the Court's scrutability project, and the majority argued in exactly this register. Its reasoning is, at bottom, an argument about comprehension: a unitary executive ensures that the public can discover "with facility and clearness the misconduct of the persons they trust," where plural or insulated administration "tends to conceal faults and destroy responsibility."[183] "When power is exercised well, the people know whom to thank; when power is exercised poorly, they know whom to blame—and whom to fire."[184] This is a responsibility scrutability intervention in clear terms. Independent agencies could operate independently when overseers could make those determinations. As administration grew inscrutable, that verification failed, and with it, in the Court's eyes, the warrant for insulation. Of course, the potential increase in scrutability bought by concentrating power and responsibility in the President may not completely rebuild accountability. As Justice Gorsuch observed in concurrence, "the fourth branch's powers still exist; they have just been reassigned to the President,"[185] and "while electoral accountability is a good thing, it cannot be the only thing."[186] Can and will the public really track the many actions of the elaborated bureaucracy, recovering decisional scrutability, even if nominally all now traceable under responsibility scrutability to the

[179] 295 U.S. 602 (1935).

[180] Trump v. Slaughter, No. 25-332, slip op. at 20–21 (U.S. June 29, 2026).

[181] *Id.* at 21.

[182] *Id.* at 27–28.

[183] *Id.* at 7 (quoting The Federalist No. 70, at 427).

[184] *Id.* at 25.

[185] *Id.* at 11–12 (Gorsuch, J., concurring).

[186] *Id.* at 12 (Gorsuch, J., concurring).

President? If the warrant for insulated expertise failed for want of verification, the question this Article takes up in Part III is whether verification can be rebuilt.

This project, however, is not one of undifferentiated hostility to agencies, and the framework developed here explains a pattern that "anti-administrativism"[187] alone does not. Where agency discretion promises to *reduce* procedural complexity, the same Court has deferred. In *Seven County Infrastructure Coalition v. Eagle County*, it cabined NEPA review, directing "substantial judicial deference" to agencies' choices about the scope of environmental analysis and condemning the litigation culture that had turned a "purely procedural statute" into an instrument of delay.[188] In *FCC v. Consumers' Research*, it declined the invitation to revive the nondelegation doctrine, sustaining a broad delegation under the traditional intelligible-principle test.[189] A Court bent on dismantling administration would have done neither, but a Court aimed at making the administrative system scrutable would retrench where complexity undermines oversight, as in *Loper Bright*, *Jarkesy*, and *Slaughter*, while deferring where deference cuts procedural sediment or where the delegation remains comprehensible and bounded.

Taken together, these decisions represent a coherent project. First, they seek to restore oversight by making administration scrutable to the courts. Then, through the President, to the public, even if doing so comes at some cost of administrative capability.[190] If legitimacy requires judicial review but agencies are acting in ways incomprehensible to the courts, then, the Court's implicit logic runs, agencies should be cut back in ways that make them susceptible to oversight. This is a plausible claim to make. The complexity crisis is real, and accountability has eroded. But the Court's remedy remains caught within the capability-accountability trap, and so creates as much of a problem as it seeks to solve. Shifting interpretive authority from agencies to courts does not make technical questions less technical, it merely relocates them to institutions less equipped to resolve them.[191] Requiring clear congressional statements on major questions does not make Congress more capable of anticipating regulatory challenges, it just creates risks that novel problems go unaddressed until political coalitions can form around specific statutory language, especially given Congress's current incapacity.[192]

---

[187] See Metzger, supra note 127, at 1–3, 15 (diagnosing "anti-administrativism").

[188] 605 U.S. ___, slip op. at 5, 7 (2025).

[189] 606 U.S. ___, slip op. at 19 (2025).

[190] Of course, it may not, as arguments around energy in the executive often propose. *See, e.g.*, Kagan, *supra* note 16, at 2331–72.

The Court's retrenchment thus threatens to produce an administrative state too constrained to address emerging challenges yet still too complex for meaningful democratic oversight. Its diagnosis, that agencies have accumulated too much power and complexity and need to be cut back, mistakes symptom for cause. The problem is not that agencies are too capable, but that capability and accountability have both eroded under the pressures of substantive and procedural complexification, leaving a system that is neither efficient nor democratic.

The question, then, is whether there is another way out of the trap. Can government become both more capable and more accountable, rather than sacrificing one for the other? The traditional doctrinal tools of allocation, review, and information-forcing have reached their limits over the settlements since 1887. Part III argues that AI offers a path forward by changing the underlying relationship between capability and accountability. Where prior technologies enhanced governmental capacity while increasing opacity, AI could represent a new chance to make government scrutable, increasing both capability and accountability. Breaking the trap will require technological transformation of the kind that government has so far failed to achieve, but that AI might finally make possible.

## III. THE FOURTH SETTLEMENT: AI BEYOND THE CAPABILITY–ACCOUNTABILITY TRAP

This Part asks whether AI could change the underlying relationship of capability and accountability that has structured administrative law through cycles of technological change. Unlike prior waves of administrative technology that increased opacity at the same time as they increased capacity, AI plausibly functions as cognitive infrastructure that helps increase *scrutability*, the comprehensibility of governance to overseers and the public. If deployed correctly, existing AI systems can translate technical material into accessible forms, surface the assumptions and tradeoffs that matter for oversight, and reduce the costs of participation and comprehension for courts, Congress, and the public. More ambitiously, advances in alignment and interpretability suggest a path toward *auditable* administration, oversight that targets the reasons and constraints that actually drive governmental determinations, rather than relying exclusively on procedural proxies that approximate substantive review.[193] The argument that follows is not a prediction that transformation is guaranteed, because politics will have its due, but aims to suggest how AI could help construct a new settlement that makes government both more capable and more accountable in the future.

---

[191] Vermeule, *supra* note 11, at 624–26.

[192] *See* Jody Freeman & David B. Spence, *Old Statutes, New Problems*, 163 U. PA. L. REV. 1, 20–22 (2014) (arguing that reliance on outdated statutes to tackle new problems through agencies is rational given congressional incapacity).

The discussion proceeds in four steps. Subpart A lays out the cognitive foundations of administrative behavior, explaining why human bounded rationality has repeatedly forced the substitution of procedure for substance, and identifies where AI could change these conditions. Subpart B argues that AI can improve public, judicial, and legislative participation and oversight by acting as scrutability infrastructure. Subpart C proposes accountability mechanisms suited to AI-assisted administration, including a Model and System Dossier that operates as an expanded administrative record for auditing governmental uses of AI, a review posture that incentivizes verifiable auditing rather than procedural completeness, and an institutional architecture for ongoing oversight of AI systems in government. And Subpart D suggests that there might be a path beyond procedural proxies, using AI to get back to something like substantive review again.

### *A. The Cognitive Infrastructure of Administration*

---

[193] I mean auditing in the expansive sense of technically-supported investigation suggested by Kroll et al., though also acknowledge the limits they point out. *See* Joshua A. Kroll et al., *Accountable Algorithms*, 165 U. PA. L. REV. 633, 660–61 (2017).

Administrative organizations are, at their core, systems for coordinating people to make decisions aimed at solving social problems under conditions of uncertainty.[194] They gather information about problems facing their constituencies, identify possible responses, model the consequences of each, incorporate preferences from the public about which consequences are desirable, and act. These tasks are performed by human minds operating within institutional structures of rules that coordinate behavior, hierarchies that allocate authority,[195] and technologies that extend cognitive capacities beyond what any individual could achieve alone.[196] The Social Security Administration, for example, combines legal and operational rules that structure it, officials who decide and act within it, and technologies, from punch cards to computers, that extend human memory and search to national scale.

Herbert Simon's model of administrative behavior provides the theoretical foundation for understanding why this "cognitive" infrastructure of administration matters. Administrators work under conditions of bounded rationality, with limited cognitive computational resources,[197] imperfect information,[198] and time pressures that push them to "satisfice" across goals rather than perfectly maximize them.[199] They cannot examine every possible course of action, perfectly predict consequences, or hold in mind the full complexity of the systems they seek to govern.[200]

---

[194] Simon, *supra* note 12, at 18–19.

[195] *See* Herbert A. Simon, *Organizations and Markets*, 5 J. ECON. PERSP. 25, 39–40 (1991) (explaining that a major use of authority is to coordinate behavior by promulgating "standards and rules of the road," enabling actors to form "more stable expectations" about others' behavior).

[196] *See* Herbert A. Simon, *A Behavioral Model of Rational Choice*, 69 Q. J. ECON. 99, 101 (1955); Simon, *supra* note 12, at 25.

[197] *See* George A. Miller, *The Magical Number Seven, Plus or Minus Two: Some Limits on Our Capacity for Processing Information*, 63 PSYCHOL. REV. 81, 81–83 (1956) (classic statement of short-term/working-memory capacity limits). *See also* Alan Baddeley, *Working Memory*, 255 SCIENCE 556, 556–57 (1992) (surveying working memory as a limited-capacity system supporting storage and processing).

[198] *See* JOHN W. PAYNE, JAMES R. BETTMAN, & ERIC J. JOHNSON, THE ADAPTIVE DECISION MAKER 1–5 (1993) (arguing decision strategies are adaptive responses of a limited-capacity information processor to task complexity).

[199] Simon, *supra* note 12, at 110–11.

[200] *See* Herbert A. Simon, *The Architecture of Complexity*, 106 PROC. AM. PHIL. SOC'Y 467 (1962) (explaining why inferring system-level properties from interacting parts is nontrivial in complex systems); Simon, *supra* note 12, at 94 ("Rationality implies a complete, and unattainable, knowledge of the exact consequences of each choice.").

Fundamentally, as Simon writes, "human rational behavior... is shaped by a scissors whose blades are the structure of task environments and the computational capabilities of the actor."[201]

Organizations help people overcome some of the limitations of bounded rationality by creating stable expectations about the behavior of other members in the organization, establishing channels of communication that facilitate coordination, and allowing specialization of work in expectation of later combination.[202] Each of these operations increases the total computational capabilities of the administrative organization. Technologies like filing, email, and statistical analysis enable people working in organizations to extend their cognitive capacities in different ways and combine with each other to become more effective.[203] An EPA scientist modeling particulate matter dispersion need not also understand the legal constraints on agency action because the lawyers handle that. A Social Security claims examiner need not design eligibility criteria because the rulemakers already did.

But specialization creates its own problems, particularly regarding oversight and control within an organization. The same division of labor that enables expertise also fragments knowledge across the organization, as scientists, economists, lawyers, and managers struggle towards mutual comprehensibility. Coordination mechanisms like meetings, memos, and hierarchical review consume resources and introduce delays. And each person working in an administrative organization has an idiosyncratic set of goals that only imperfectly corresponds to the goals of the organization, which they pursue sometimes at the expense of the organization's goals.[204] Organizations have tools for aligning individual to institutional goals, but these alignment and oversight mechanisms are themselves costly to run, as managers must spend resources to monitor subordinates, provide incentives, and cultivate loyalty.[205] And oversight costs rise with specialization and volume. Where an official or agency is expert enough, substantive oversight of her decisions becomes impossible even for her own managers.[206] These risks run highest in agencies, whose purpose is precisely specialized, expert problem-solving.[207]

---

[201] Herbert A. Simon, *Invariants of Human Behavior*, 41 ANN. REV. PSYCHOL. 1, 7 (1991).

[202] *Id.* at 19. Simon, *supra* note 12, at 39–40.

[203] *See* Lee Sproull & Sara Kiesler, *Reducing Social Context Cues: Electronic Mail in Organizational Communication*, 32 MGMT. SCI. 1492, 1492–94 (1986) (showing email alters organizational information exchange and coordination, not merely speed).

[204] *See* Michael C. Jensen & William H. Meckling, *Theory of the Firm: Managerial Behavior, Agency Costs and Ownership Structure*, 3 J. FIN. ECON. 305, 308–10 (1976) (defining agency relationships and why agents' incentives diverge from principals').

These dynamics are internal to the administrative organization. But the capability-accountability trap arises from a further problem: the relationship between the organization and its external overseers necessary to create democratic legitimacy in government. Courts, Congress, and the public must evaluate whether agencies are acting lawfully, effectively, and in the public interest. To do so, they need to understand what agencies are doing and why. This understanding is what I call *scrutability*: the cognitive tractability of administrative action for those who did not participate in producing it.

Scrutability is constrained by the same bounded rationality that constrains administration itself. Judges have limited time, small staffs, and no specialized training in the domains they review. Members of Congress face attention scarcity across hundreds of policy areas and rely on skeletal committee staff. Citizens have lives to live and cannot devote significant cognitive resources to monitoring agency behavior. When administrative action is simple, that is when the relevant considerations are few, the causal chains short, and the tradeoffs intuitive, external overseers can evaluate substance directly, like a nineteenth century court assessing whether a municipality had authority to grant a particular license.

However, when administrative action grows complex, scrutability declines. The shift from simple licensing to railroad rate-setting, from local poor relief to national social insurance, and from common law nuisance to probabilistic environmental risk assessment each represented an increase in the sophistication required to assess whether an agency had acted correctly that soon exceeded the capacity of the public or appointed overseers. Over time, through increasing scale and scientific sophistication administration had become substantively inscrutable.

[205] Simon, *supra* note 12, at 9.

[206] Jensen & Meckling, *supra* note 204, at 308–10 (agency costs include monitoring, bonding, and residual loss).

[207] Stephenson, *supra* note **Error! Bookmark not defined.**, at 53–56.

As Parts I and II traced, oversight compensated by moving from the substantive to the procedural grounds where generalist overseers are cognitively equipped. While they cannot make advanced scientific determinations, courts can examine records, reasons, responses to objections, and due process requirements, while Congress lays out procedure and deputizes affected parties to sound fire alarms. Modern administrative law effects this procedural turn. The administrative record focuses review on what can be compiled and transmitted to a generalist overseer; *Chenery* requires agencies to act based on the reasons in that initial record; and "hard look" review centers oversight on whether the agency confronted relevant material and explained its choices. Information-forcing regimes like FOIA, the Sunshine Act, and NEPA create accountability through informational documentation and disclosure. This architecture does produce constraint, but it incentivizes agencies to invest in review-proof paper trails rather than directly better decisions.[208] Over time, procedural layers compound into a scrutability tax that crowds out substance, privileges repeat players experienced in handling procedural complexity, and makes engaging with the lengthy record impossible for parties with limited resources.[209] The structure of government becomes inscrutable alongside its content.

These tradeoffs operate differently across the administrative state. Broadly, there are three types of administrative action, distinguished by the nature of the decisions involved and the role of technical expertise:

*Type 1: High-volume, low-discretion adjudication*. Consider benefits determinations, licensing decisions, permit approvals, and other administrative adjudications that involve applying relatively determinate criteria to individual cases. The administrative function in these domains is matching facts to legal categories: Does this applicant's medical condition meet the statutory definition of disability? Does this invention satisfy the non-obviousness requirement? Here are the rules for each. The goal is predictable, consistent, accurate, and fair application, and the discretion of officials is limited to advance those objectives.

---

[208] Bagley, *supra* note 108, at 351–52.

[209] *See* Wendy E. Wagner, *Administrative Law, Filter Failure, and Information Capture*, 59 DUKE L.J. 1321, 1324–28 (2010).

Procedural review can operate here. The criteria used in Type 1 determinations are articulable *ex ante* and often based in statute, so reviewing courts can assess whether the agency applied the right standard and gave required due process,[210] even if they cannot evaluate the underlying factual determinations with confidence in every case appealed to them. The problem is that such appellate review is limited by the sheer volume of cases. Statistical evaluation could improve accountability. Mass adjudication is a statistical system with error and bias rates that could be estimated through sampling and appeals, and measures implemented to improve them. Generative AI could also improve this Type of adjudication, for example by generating personalized explanations of why an applicant is denied welfare benefits.

*Type 2: Technical rulemaking with complex records*. Rules like EPA air quality standards and FDA drug approvals require synthesizing scientific evidence, economic analysis, and policy judgments into generally applicable requirements. The record in a major rulemaking is a complex artifact integrating multiple disciplinary perspectives in which courts lack expertise. The agency must characterize the problem (what harms does the pollutant cause?), evaluate evidence (what does the epidemiology show?), model consequences (how will affected parties respond to different regulatory options?), and make tradeoffs (how much cost is justified to achieve how much risk reduction?). While Congress sometimes guides agencies in making these determinations, often there is little specific statutory guidance in how they should be done, and indeed outsourcing the details to experts is the point of delegation.

Procedural review faces difficulties in Type 2 contexts. Courts can check that the agency compiled a record and offered reasons, and sometimes even look to the quality of these reasons, but they cannot evaluate whether a key scientific conclusion was substantively sound. Sophisticated challengers learn to exploit the focus on procedure and box-checking that this dynamic generates, flooding the record with studies and objections to create procedural litigation hooks. Agencies respond in the "defensive crouch", with the effect of extending rulemaking timelines and increasing costs without necessarily improving outcomes.[211]

[210] *See* Henry J. Friendly, *Some Kind of Hearing*, 123 U. PA. L. REV. 1267, 1277–1304 (1975).

*Type 3: High-salience, value-laden tradeoffs*. Some administrative decisions involve fundamental normative choices about which reasonable people disagree, even where there is scientific consensus. How should we balance tradeoffs between reducing emissions and disrupting economic production? What balance should be struck between drug safety and patient access? How should agencies weigh privacy against security in surveillance contexts? Politicians debate these questions in elections and in Congress, but often it is agencies that make the key determinations in their application in the world. These questions have technical dimensions, but their resolution ultimately depends on value judgments that science cannot dictate.[212]

Procedural review is least adequate for Type 3 decisions. The major questions doctrine reflects the sense among courts that some choices are too important and political for agencies to decide them without clear congressional delegation. But Congress often cannot or will not specify answers to these questions, because it may lack the technical capability to resolve them, because they may emerge after legislation is already passed, or because no new law can get through congressional gridlock. Agencies step in to resolve critical societal problems like climate change without clear authority to make the relevant value judgments, and procedural review cannot supply the fundamental democratic legitimacy that congressional authorization would provide.

This typology clarifies where AI might help and where it cannot. The scrutability problem is most tractable in Type 1 adjudications, where AI systems can match facts to criteria using their natural language capabilities and the main difficulties are scale and documentation. Indeed, AI can likely improve Type 1 adjudications significantly, both for regulated parties and for agencies. AI systems that can process applications, flag edge cases, and generate individualized explanations could dramatically reduce the scrutability tax while maintaining or improving accuracy. The challenge is ensuring that AI-assisted adjudication satisfies due process requirements, most significantly that affected parties can understand and contest the grounds for decisions affecting them, but having AI assistance would likely make these adjudications easier to understand.

[211] *See* Bagley, *supra* note 108, at 351–52.

[212] *See* Jasanoff, *supra* note 12, at 1–32.

Type 2 contexts present more substantial challenges. AI cannot resolve scientific uncertainty or make contested empirical questions disappear, but it can reduce the costs of producing and processing the complex records that technical rulemakings generate. Advanced AI is starting to be used in different scientific domains to answer empirical questions, and plugging scientific models into LLM-based frontier AI systems that can model regulatory domains, summarize generated evidence, and translate technical findings into language that is comprehensible to agencies, their overseers, and the public could improve scrutability without pretending to resolve substantive disputes. The goal is not to replace expert judgment but to make that judgment more auditable, agencies benefiting from the capabilities of this advancing technology.

Type 3 contexts present the most serious risks from AI. Value-laden tradeoffs require democratic legitimation that technical systems, no matter how capable, cannot supply. This principle is core to the purpose of judicial oversight of the administrative state. AI systems that purport to resolve such tradeoffs without human participation risk "values laundering," obscuring normative choices behind a veneer of algorithmic neutrality.[213] In major questions-type cases, AI should not make decisions but rather clarify what decisions are being made. These systems could surface tradeoffs, identify winners and losers from different policy alternatives, and present the choice in terms that democratic institutions can evaluate. AI systems must be designed to prevent AI-assisted framing from manipulating rather than informing deliberation, but there is a chance to significantly improve how administration handles this type of problem here.

More generally, AI might enable oversight to depend less on procedure and more on substance or even true alignment of agencies to the public will, reducing procedure's attendant ossification. These systems can understand the world and perform research, identifying and using relevant information and demonstrating advanced knowledge of scientific domains critical for modern governance both in and outside the lab.[214] Multimodal systems can take in data of various forms and use coding tools to analyze it statistically, finding important patterns.[215] When deployed in agentic environments, these systems can identify problems and opportunities, and sometimes even solve them. Courts and Congress could ask AI systems about a given agency decision, creating a new means for low-cost provision of expertise that enables the overseeing bodies to at least somewhat substantively evaluate whether an agency acted correctly. Furthermore, advances in technical alignment, including reinforcement learning from human feedback, constitutional AI methods, and related techniques, can bring AI systems' objectives closer to human ones, while interpretability might allow deeper examinations of AI reasoning than is possible for human cognition. Truly aligning AI systems to the public good would mean that oversight of AI decisions is less necessary, reducing its social costs.

---

[213] *See* Andrew D. Selbst et al., *Fairness and Abstraction in Sociotechnical Systems*, *in* FAT* '19: Proceedings of the Conference on Fairness, Accountability, and Transparency 59, 59–68 (2019). Though note that the process of scientific advising for political questions itself has been described as a kind of values laundering through the apparent neutrality of science. *See* Jasanoff, *supra* note 12, at 1–32.

The next two Subparts develop these possibilities. Subpart B examines how AI can serve as scrutability infrastructure for participation and oversight—reducing the costs of understanding agency action and enabling more meaningful engagement by courts, Congress, and the public. Subpart C proposes accountability mechanisms suited to AI-assisted administration, including an expanded administrative record, review postures that reward transparency, and institutional architectures that support ongoing audits of agency actions. Together, they sketch a candidate settlement that uses AI to restore substantive accountability rather than merely adding new procedural layers to the accumulated sediment of past reforms, seeking to provide a positive vision for how to transform administration in this forthcoming technological era.

---

[214] *See* Karan Singhal et al., *Large Language Models Encode Clinical Knowledge*, 620 NATURE 172 (2023) (evaluating LLM clinical knowledge); Bernardino Romera-Paredes et al., *Mathematical Discoveries from Program Search with Large Language Models*, 625 NATURE 468 (2024) (FunSearch discovers new algorithms/scientific results); Daniel J. Mankowitz et al., *Faster Sorting Algorithms Discovered Using Deep Reinforcement Learning*, 618 NATURE 257 (2023) (AlphaDev discovers improved sorting routines integrated into LLVM).

[215] *See* OPENAI, GPT-4 TECHNICAL REPORT 1 (2023) (multimodal inputs).

### *B. AI as Scrutability Infrastructure for Participation and Oversight*

The procedural substitution described in Subpart A was a reasonable response to declining scrutability resulting from the increased cognitive costs of handling the complexity of government. If overseers cannot evaluate substance because they lack the time and resources to develop relevant expertise, they can at least police process. But procedural proxies have limits. Today, overseers often verify that agencies followed required steps without confirming that those steps produced good outcomes, creating compliance burdens that accumulate over time, and ultimately advantaging sophisticated parties able to bring substantive and procedural expertise to bear over diffuse publics that can't.

Done right, AI offers a different approach. Rather than substituting procedure for substance, use technology to restore substantive comprehension and create more accountable but also acceptably complex government. If the scrutability problem arises because complex agency action exceeds the cognitive capacity of overseers and participants, then tools that reduce the cognitive costs of understanding or that increase that effective cognitive capacity could ameliorate the problem at its source.[216] In this view, AI serves as a kind of scrutability infrastructure that does not replace human judgment, but enables humans to exercise their judgment over material they could not otherwise comprehend without significant investment.

This Subpart examines three contexts where AI-enhanced scrutability could transform administrative governance: rulemaking, where AI can improve agency information-gathering, modeling, action, and response, and reduce barriers to meaningful public participation; adjudication, where AI can improve both the accuracy of initial determinations and the accessibility of review to regulated parties; and congressional oversight, where AI can help legislators focus their scarce attention on agency choices relevant to their actions. Each of these proposals is an intervention on decisional scrutability, reducing the cost, for participants and overseers alike, of understanding what an agency decided and why. This Subpart then addresses the risks, including manipulation, astroturfing, and inequality in the use of technologies, that AI-assisted participation creates and proposes initial safeguards to mitigate them, though accountability is discussed in detail in the following Subpart.

[216] *See* Andy Clark & David Chalmers, *The Extended Mind*, 58 ANALYSIS 7, 7–12 (1998).

1. Rulemaking: From Mass Comments to Structured Participation

Notice-and-comment rulemaking is the paradigmatic form of public engagement in administration. The APA requires agencies to publish proposed rules, accept written comments from interested parties, and respond to significant objections before issuing final rules.[217] In theory, this process enables democratic input into technical policymaking because affected parties can inform agencies about real-world impacts, identify errors in analysis, and propose superior alternatives to those the agencies have put forward.[218]

In practice, notice-and-comment often fails to achieve these goals. Major rulemakings generate thousands or millions of comments, most of which are duplicative form letters that provide little contextual information.[219] Agencies cannot meaningfully handle this volume of comments, so they triage, prioritizing comments from sophisticated parties who frame objections in terms that threaten lawsuits.[220] At the same time, ordinary people affected by a proposed rule often lack the knowledge that the rule exists and don't have the ability to understand the proposal or how it might affect them. Notice-and-comment is thus dominated by repeat players who can invest in technical engagement rather than operating as a true form of democratic input.

AI can help address these problems. Consider a layered AI interface that enables citizen participation while preserving the procedural protections of the APA:

*Layer 1: Record-anchored explanation.* Alongside the Notice of Proposed Rulemaking, agencies would publish explanatory materials generated by AI from agency research and the record that put the proposed rule in terms that normal citizens could understand. An agency could also deploy a chatbot to help people process those materials. AI analysis would identify what the rule changes, who is affected, what tradeoffs the agency is making, and what alternatives were considered, and the chatbot could then answer questions based on the particular circumstances of affected parties that use it. The chatbot's explanations would be anchored to the administrative record, citing the relevant underlying technical document, data source, or analytical assumption. This layer is achievable with current technology. LLMs can summarize complex documents, answer questions about their contents, and maintain citation links to source material.[221] Hallucination and similar risks, and their mitigation, are considered below.

---

[217] 5 U.S.C. § 553(b)–(c).

[218] *See* Stewart, *supra* note 130, at 1711–60.

[219] *See* Mendelson, *supra* note 156, at 1346–50, 1359–76.

[220] *Id.*

This layer should not create significant new procedural costs, and agencies could work with frontier AI providers to develop their chatbot aids on top of powerful existing tools. Chatbot explanations should be legally non-binding, aids to understanding the record but not themselves authoritative interpretations.[222] The agency would still be obligated to respond to comments on the actual proposed rule, with the AI layer simply reducing the cost of understanding that rule for affected parties.

*Layer 2: Structured commenting.* Beyond explaining what's in the record, AI could help commenters articulate their views in forms that agencies would pay attention to and process. Current notice-and-comment interfaces accept free-text submissions, creating an aggregation problem: how should an agency weigh a heartfelt but naïve letter from a layperson against a sophisticated critique produced on behalf of a sophisticated party by sophisticated counsel? How much weight should a mass of identical form letters get?

An AI-powered structured commenting interface would help lay citizens translate their concerns into more effective forms. After helping users understand the proposed rule and its underpinnings, the system could guide them in identifying how they are affected, what they dispute about it, what evidence they can muster in support of their position, what alternative they might prefer, and what tradeoffs they might be willing to accept. The AI would help users, who otherwise might lack technical knowledge, understand what they are agreeing to or contesting, identify the parts of the record relevant to their concerns, and frame their comments in terms the agency must address. The contextual knowledge that notice-and-comment is partly supposed to elicit would be better expressed through this kind of aid. Where the commenter does not actually have a germane objection, the system could help them understand that too.

---

[221] *See* Paul Rust et al., *Evaluation of a large language model to simplify discharge summaries and provide cardiological lifestyle recommendations*, 5 COMMS. MED. 208, 208 (2025).

[222] *See* 5 U.S.C. § 553(b)(A).

This is more ambitious than Layer 1, requiring not just summarization but interactive dialogue that maps user concerns onto the rulemaking. It also raises design questions about the affordances and performance of the AI system. How much should the system shape user input? Is there a risk that structured commenting would force diverse concerns into pre-approved categories that miss nuance? These are real issues, but the system still represents an improvement on the status quo. And it could also be designed such that unstructured input remains available for those who want it.

*Layer 3: Comment aggregation and transparency.* Layers 1 and 2 have focused on how AI can improve the experience of parties affected by a rulemaking, but agencies also stand to benefit from the use of advanced AI. Masses of comments on important rules (remember the 700,000 from the EPA's air quality standard rulemaking) already force agencies to triage and prioritize among the comments received. Agencies must develop heuristics based on substantive factors and the risk of procedure-based lawsuits to determine things like which comments receive substantive responses, how duplicates are handled, what weight attaches to different types of input, and the like. AI can make this process more systematic and transparent.

AI-powered comment clustering[223] would group submissions by the issues they raise, creating a map of public concerns that agencies must address. The agency's responses could be organized around clusters rather than individual submissions, explaining how each significant objection described across a set was considered. Different algorithmic weights could be given for different kinds of comments, and those weights disclosed. For example, duplicative comments from form letters could be counted for prevalence (indicating public salience) without being treated as independent technical evidence. Near-duplicate comments generated by AI systems and submitted by single entities attempting to influence the rulemaking could be flagged and handled appropriately.

---

[223] *See generally* BRIAN S. EVERITT, SABINE LANDAU, & MORVEN LEESE, CLUSTER ANALYSIS (2009).

This layer helps address ossification. If agencies can process public input more efficiently by using AI to identify sets of substantive objections and document how each was addressed, the burden of notice-and-comment on agency staff declines. At the same time, the agency maintains an equal or greater understanding of the public input, relying on powerful AI rather than human summarization. Faster rulemaking becomes possible without sacrificing the democratic input that notice-and-comment is supposed to provide.

2. Adjudication: Procedural Justice at Scale

Administrative adjudication presents different challenges than rulemaking. Where rulemaking produces generally applicable standards, adjudication applies standards to individual cases. The due process stakes are direct because they affect things like the disability benefits or immigration status of specific people. Procedural justice requires that affected parties understand the grounds for decisions affecting them and have meaningful opportunity to contest adverse determinations.[224]

However, high-volume adjudication has long put serious strains on due process and the values that it serves. The Social Security Administration decides millions of disability claims annually.[225] Immigration courts face crushing backlogs, leading to serious delays.[226] Agencies across federal and state governments make countless determinations that affected individuals may never fully understand.[227] Take a form letter denying a claim, often the only explanation an adversely affected party will get. It is boilerplate, citing obscure regulatory criteria, and perhaps satisfying minimal legal due process requirements but providing little actual comprehension for the affected party. Claimants often do not know why they were denied, what evidence they would need to marshal to support their case, or how to navigate the appeals process. Putting the burden of understanding this process on regulated parties reduces the burden on government and allows it to continue to operate at scale, but it does not serve justice.

[224] *See* Friendly, *supra* note 210, at 1277–1304.

[225] SOC. SEC. ADMIN., FY 2024 ACTUAL PERFORMANCE: WORKLOAD AND OUTCOME MEASURES 1 (2024).

[226] Holly Straut-Eppsteiner, *FY2024 EOIR Immigration Court Data: Caseloads and the Pending Cases Backlog* 1, 4, CONG. RSCH. SERV. (IN12492, Jan. 24, 2025).

AI could improve procedural justice in adjudication across three dimensions: the quality of initial determinations, the comprehensibility of decisions, and the accessibility of appeals.

*Initial determinations.* AI-assisted case processing could improve accuracy and consistency in high-volume adjudication. Most adjudications are, at their foundation, simple classification tasks of the kind that AI systems are well-equipped to handle. In the paradigm cases, a rule laying out a set of criteria is applied to an individual to make a binary decision. Take a system that uses AI as a decision aid for human adjudicators. Systems trained on past determinations could flag cases that warrant closer review, identify relevant evidence that adjudicators might have missed, and detect inconsistencies in decisions across similar cases. This is Type 1 administrative action in the typology introduced in Subpart A: high-volume, relatively low-discretion application of criteria to facts. More ambitiously, centralizing adjudication in one or a few task-specific AI systems would permit more rigorous evaluation and correction in ways human-staffed adjudication cannot. When one of thousands of human adjudicators has idiosyncratic bias, correcting a single decision on appeal does little to improve overall system performance or likely even the later performance of that person. But a judicial decision requiring improvement of an algorithm propagates across every case the system handles, making system-level oversight tractable.

---

[227] Donald Moynihan, Pamela Herd, & Hope Harvey, *Administrative Burden: Learning, Psychological, and Compliance Costs in Citizen-State Interactions*, 25 J. PUB. ADMIN. RSCH. & THEORY 43, 43–46 (2015).

Such uses of AI present real risks. Automation bias, error propagation, and opacity all undermine procedural justice in AI administration.[228] Subpart C addresses the accountability mechanisms necessary to manage these risks. Here, the point is simply that AI-assisted initial processing can improve baseline accuracy in systems where current error rates are substantial and inconsistency is endemic. Human review has deep problems, and AI can help us do better.

*Comprehensible decisions.* AI can transform the explanation of adverse determinations. Rather than boilerplate denial letters, agencies could provide individualized explanations at much lower cost than having a human write it. For example, here is a simple explanation to an affected party in a disability hearing: "Your claim was denied because the medical evidence you submitted did not establish that your condition prevents you from performing sedentary work. The vocational expert found that jobs X, Y, and Z remain available to someone with your functional limitations. To succeed on appeal, you would need to provide evidence showing [specific deficiency]." Template form letters can partially reduce the cost of personalization today, but would be far cheaper with AI.

Such a step moves us toward constitutionally grounded due process. *Mathews v. Eldridge* requires that procedures be calibrated to the stakes involved and the risk of erroneous deprivation.[229] Individualized explanation reduces error by helping claimants identify when the agency has made a mistake and by focusing appeals on actual points of dispute. It would also improve the perceived legitimacy of adverse decisions because people are more likely to accept unfavorable outcomes when they understand the reasons.

Current large language models can generate such explanations.[230] The system would need access to the case record, the criteria applied, and the specific findings that led to denial. It would need to be accurate, which requires measuring and addressing hallucinations.[231] The system would also need safeguards against manipulation from within the agency, because the explanation should reflect the actual reasons for decision, not a *post hoc* justification designed to discourage appeals. However, human-written explanations present the same problem, and doctrinal requirements that they reflect the agency's true reasoning are of limited effect because it is difficult to see into the reasoning processes of human officials.

---

[228] *See* Coglianese & Lehr, *supra* note 157, at 1213–17; Aziz Z. Huq, *A Right to a Human Decision*, 106 VA. L. REV. 611 (2020).

[229] 424 U.S. 319 (1976).

[230] *See* Rust et al., *supra* note 221.

*Accessible appeals.* The same interface that explains decisions can help claimants with appeals. What additional evidence would strengthen a claimant's case? What are the deadlines? What procedural steps must they take to launch an appeal? The AI helps claimants who cannot afford representation by human lawyers, a partial remedy for the inequality that pervades administrative adjudication.

This application raises concerns about unauthorized practice of law.[232] The system should not provide legal advice in the sense of making rights-affecting recommendations about whether and how to appeal, but instead it should provide information about process and requirements that claimants are entitled to know. The line is blurry, but it is the same line that legal aid organizations and court self-help centers must work across daily. AI could dramatically expand claimants' access to useful information and provide procedural guidance that is currently available only to those with resources to obtain representation, and the benefits of such tools should not be overlooked.

### 3. Congressional Oversight: AI as Attention Multiplier

Congress is structurally incapable of overseeing the administrative state. Members have limited time, divided across hundreds of policy areas and the demands of campaigning and constituent service.[233] Personal staffs are small, can go deep on only a few issues, and are not growing. Committee staff possess expertise in different domains but are also stretched thin, and cannot match agency specialists who combine expertise with substantial time invested in the problems at issue.[234] The "information asymmetry" between agencies and their legislative overseers is vast, and, under existing procedures, agencies mostly control the production of the records that oversight relies upon.[235]

---

[231] *See* Yue Zhang et al., *Siren's Song in the AI Ocean: A Survey on Hallucination in Large Language Models*, arXiv:2309.01219 (Sep. 14, 2025).

[232] *See Modernizing Unauthorized Practice of Law Regulations to Embrace AI-Driven Solutions and Improve Access to Justice*, AI POLICY CONSORTIUM FOR LAW & COURTS (Aug. 2025).

AI cannot solve Congress's fundamental attention scarcity or its political dysfunctions. But it can reduce the costs of informed oversight, enabling legislators to focus their limited attention on consequential agency choices that are relevant to their interests and their constituents. Consider an AI system that operates as a congressional query interface for major rulemakings. Every rule and its associated record is uploaded to some database, which an AI processes. Staff could use this AI assistant to interrogate the rulemaking record through natural language, asking about the assumptions driving a cost estimate, what alternative courses the agency considered or should have, what statutory terms the agency is interpreting and relying on, and the like. This interface would substantially reduce the cost of sitting down with a complex record, increasing effective congressional capabilities. The AI could also learn what specific Members of Congress care about and proactively flag rulemakings or patterns of agency activity that might be of interest to them.

This capability matters most for allocation of interpretive authority, the first pillar of the administrative settlements. Post-*Loper Bright*, courts will exercise independent judgment on statutory interpretation, but Congress retains primary authority to resolve ambiguity through legislation. The problem is that Congress often does not know which interpretive questions will matter until litigation surfaces them years later, and changing circumstances and new technologies can create new ambiguities in statutory language where none existed before. AI-assisted oversight could identify interpretive pressure points as they arose. For example, a system could flag to an interested legislator or her staff that a new rule interprets "stationary source" in a way that extends or shrinks the agency's authority to regulate new pollution categories or that a new guidance interprets "waters of the United States" to include new features. Rather than waiting for litigation or fire-alarm interest groups, Congress could intervene prophylactically.

---

[233] Russell W. Mills & Jennifer L. Selin, *Don't Sweat the Details! Enhancing Congressional Committee Expertise Through the Use of Detailees*, 42 LEGIS. STUD. Q. 611, 613 (2017).

[234] *Id.* at 611–13.

[235] McCubbins, Noll, & Weingast, *supra* note **Error! Bookmark not defined.**, at 250–51.

The interface could also surface the value-laden choices that are embedded in technical analysis. For example, cost-benefit analyses require assumptions about discount rates, the value of statistical lives, and the baseline against which impacts are measured.[236] Choices about what values to use encode normative judgments about how to weigh present against future harms, whose lives count, and what world we are comparing the regulatory intervention to.[237] AI systems can clarify the assumptions being made and transform them into something that legislators can engage with like a statement that "the agency's conclusion that benefits exceed costs depends on valuing statistical lives at $X, discounting future harms at Y%, and assuming baseline Z. Here is how the conclusion changes under alternative assumptions."

None of this substitutes for congressional will. If legislators do not want to oversee agencies, perhaps because oversight is politically costly, or because they prefer to blame agencies for unpopular outcomes, AI cannot compel them to do so. But if Congress wants to exercise meaningful supervision, AI could reduce its cost. It reduces the extent to which Congress must rely on proxies like fire-alarm oversight[238] toward something more systematic without requiring resources that Congress does not have.

Implementation would require the development of institutional infrastructure. Congressional support agencies, including the Government Accountability Office, the Congressional Research Service, and the Congressional Budget Office could develop and maintain AI tools for legislative oversight. A revived Office of Technology Assessment[239] or the U.S. Center for AI Standards and

---

[236] Office of Mgmt. & Budget, Exec. Office of the President, Circular A-4 (2023).

[237] *See* CASS R. SUNSTEIN, THE COST-BENEFIT STATE: THE FUTURE OF REGULATORY PROTECTION (2002).

[238] *See* McCubbins & Schwartz, *supra* note 137, at 166.

[239] *See* Darrell M. West, *It is time to restore the US Office of Technology Assessment*, BROOKINGS (Feb. 10, 2021), https://www.brookings.edu/articles/it-is-time-to-restore-the-us-office-of-technology-assessment/.

Innovation[240] could provide independent technical analysis of agency systems and actions. The investments necessary for such oversight are modest compared to the significance of effective congressional oversight of AI administration.

4. Safeguards: Propaganda, Astroturfing, Inequality, and Capture

AI-enhanced participation creates new risks that require new safeguards. Agencies could use AI to act in ways contrary to the public interest, turning the interfaces intended to translate complexity into publicly-comprehensible terms into tools that manipulate rather than inform. External groups could also use AI to try to influence or capture agencies, building on existing AI astroturfing efforts.

Four categories of risk require particular attention:

*Propaganda risk.* Under the model of AI in government just discussed, agencies would control the AI interfaces that explain their proposals. Measures would need to be taken to prevent an agency from designing an AI to favor its preferred outcome, for example by emphasizing benefits, downplaying costs, or framing alternatives uncharitably. Agencies would not need to lie outright but instead deploy current rhetorical strategies like selective presentation in preambles and fact sheets, now with technological augmentation the imprimatur of technological neutrality.

Mitigation could include structural constraints on interface design. Agencies should be legally required to present credible alternatives and their strongest supporting arguments. Review of interface outputs by OIRA or an independent AI-expert body could verify that explanations are balanced through sampling-based substantive review. If a party identified that an agency had violated these principles, that could be independent grounds for a suit to reverse the relevant decision. Interface design choices might be made part of the record and associated evidence for an agency action, enabling affected parties to identify and challenge bias.

[240] *See Statement from U.S. Secretary of Commerce Howard Lutnick on Transforming the U.S. AI Safety Institute into the Pro-Innovation, Pro-Science U.S. Center for AI Standards and Innovation*, U.S. DEPT. COMMERCE (Jun. 3, 2025), https://www.commerce.gov/news/press-releases/2025/06/statement-us-secretary-commerce-howard-lutnick-transforming-us-ai.

*Astroturf risk.* Rulemakings are already being flooded by AI-generated synthetic comments.[241] The comment process will become even noisier, and agencies could soon face an impossible task in distinguishing real from manufactured public input.

Mitigation requires authentication and provenance mechanisms. Identity verification (at least to the level of confirming that a comment comes from a real person) could filter out purely synthetic submissions. AI watermarking or evaluations of whether the text is human- or AI-written could enable appropriate weighting of different kinds of comments without excluding AI-assisted comments entirely, particularly for reasons of fairness and accessibility.[242] These detection techniques are still nascent but represent the beginnings of a way to raise the costs of astroturfing.

*Inequality risk.* AI tools are not equally accessible. Sophisticated parties will be able to spend more on better systems and on inference, fine-tuning their use of AI interfaces, optimizing prompts, and adversarially exploiting government capabilities in ways that unsophisticated users cannot match. AI-enhanced participation could reproduce existing access inequalities rather than ameliorating them.

This risk is real but should be compared to the current domination of participation by sophisticated parties, discussed in Part II above. Repeat players might benefit from AI, but they are starting from a much higher baseline, so the marginal improvement might still go to ordinary participants. The goal is to ensure that citizens without resources can participate meaningfully, even if participation remains easier for those with resources. Government-provided AI interfaces, like those in Part III.B.1-2, should be optimized for accessibility, focusing on plain language, intuitive navigation, and responsive assistance for confused users. Public investment in user-friendly tools could partially offset private investment in optimization.

---

[241] Mark Febrizio, *Will ChatGPT Break Notice and Comment for Regulations?*, GEO. WASH. UNIV. REG. STUD. CTR. (Jan. 30, 2023), https://regulatorystudies.columbian.gwu.edu/will-chatgpt-break-notice-and-comment-regulations.

[242] *See* Aloni Cohen, Alexander Hoover, & Gabe Schoenbach, *Watermarking Language Models for Many Adaptive Users*, arXiv:2405.11109 (Jun. 28, 2024).

*Infrastructural capture risk.* Who builds and maintains the AI systems that agencies use? If vendors develop and control the infrastructure of administrative participation, they could acquire leverage over the agencies that depend on them.

Mitigation requires attention to procurement and governance. In particular, audit rights, portability, in-house evaluative capacity, and, for core functions, public development of AI infrastructure, would be steps in the right direction. Part III.C develops these prescriptions in the accountability context, but they also matter at the infrastructural level.

Taken together, the safeguards described above would help legitimate AI-assisted administration. AI-enhanced scrutability depends on design choices that prioritize public comprehension and democratic access. Subpart C turns from enabling participation to ensuring accountability, and thus directly to administrative law. It lays out a proposal for how administrative law should adapt its mechanisms of record, review, and information-forcing to a world where AI systems play significant roles in agency decision-making to ensure that AI-assisted administration remains subject to meaningful oversight.

### *C. AI as Accountability Infrastructure*

Subpart B described how AI can improve scrutability for those who participate in and oversee administrative governance. But participation and oversight require something to oversee. The administrative record, developed across the settlements examined in Part I, serves this function for human-driven administration. Agencies, when acting, must compile documentation of the information considered, the alternatives evaluated, and the reasons for their choices, which can provide the basis for public response, oversight, and lawsuits. Courts can review this record to determine whether agencies have met the requirements of rationality and legality.

AI-assisted administration requires a new kind of record. When a frontier AI makes a decision, the "reasons" for that decision exist, if anywhere, in activation patterns that humans are presently unable to interpret and that are based on weights and masses of training data that combine in complex forms. If accountability depends on comprehensible records, and AI reasoning is not comprehensible in traditional forms, then accountability requires a new form of record.

This Subpart proposes an accountability infrastructure that would enable AI-assisted administration by making it susceptible to oversight. It begins by describing how AI can help relocate complexity so that government retains the sophistication to solve complex problems while also making it reviewable and auditable. It then proposes a doctrinal framework that, combined with technical investment in auditing and verification, would make AI reasoning reviewable.

### 1. The Relocation of Complexity

Before diving into the doctrinal framework, it is worth noting that the AI administration proposed here does not seek to simplify administration as a whole but rather shifts complexity from one part of government to another. Complexity is bad where it overwhelms the humans who have to engage with it. AI can simplify the interface through which courts, Congress, and the public interact with administration while adding useful backend complexity in the form of model architectures, training pipelines, evaluation procedures, vendor relationships, security considerations, drift monitoring, and associated practices that improve both governmental capability and reviewability.

Relocating complexity is necessary because modern, scientific problems require complex government. However, complexity has ended up in the wrong places, focused in incomprehensible substantive determinations and accumulated procedure. AI addresses this by shifting complexity to systems that can be audited through technical means, while presenting simplified interfaces to humans who cannot handle the underlying detail. A court cannot evaluate whether an atmospheric dispersion model uses appropriate parameters, but it could evaluate whether an agency followed appropriate procedures for validating the model, testing its performance, and monitoring its behavior over time. Such an approach leverages the strengths of procedural review while updating it for the AI era, using AI's technical characteristics to make that review more efficient and so less ossifying.

The legal goal, then, is not to eliminate complexity but to make it auditable. Auditability means that the behavior of AI systems can be evaluated and verified by parties with appropriate expertise,[243] operating on behalf of the courts or legislators who ultimately hold agencies accountable. Courts verify that agencies followed appropriate audit procedures while auditors verify that AI systems substantively behave as claimed. This combination provides assurance that AI-assisted decisions are lawful and reasonable. The chain also restores responsibility scrutability because it helps isolate where failures in review have occurred.

### 2. The AI Administrative Record: A Model and System Dossier

Adapting administrative accountability for AI requires an expanded administrative record that documents how AI systems are designed, tested, deployed, and monitored. This "Model and System Dossier" would be a required addendum to the traditional administrative record. It would be required whenever an agency adopts a new AI system that it will rely on in rulemaking or adjudication. The Dossier, inspired by existing frameworks like OMB's government-wide guidance on agency AI use,[244] the EU AI Act and its Code of Practice,[245] the Frontier Safety

---

[243] *See* Edwin A. Farley and Christian R. Lansang, *AI Auditing: First Steps Towards the Effective Regulation of Artificial Intelligence Systems*, 38 HARV. J. L. & TECH. DIGEST (Feb. 19, 2025), https://jolt.law.harvard.edu/digest/ai-auditing-first-steps-towards-the-effective-regulation-of-artificial-intelligence-systems.

[244] Office of Mgmt. & Budget, Exec. Office of the President, *2025 Federal Agency Artificial Intelligence Use Case Inventory* (updated Apr. 13, 2026).

[245] European Commission, *General-Purpose AI Code of Practice* (July 10, 2025), https://digital-strategy.ec.europa.eu/en/policies/contents-code-gpai.

Frameworks of leading AI companies,[246] and model cards[247] and datasheets,[248] but adapted for administrative law, could include the following elements, among others:

*System identity, scope, and decision role.* What is the AI system, and what is it used for? This element defines the boundaries of the system subject to documentation and distinguishes consequential AI applications from trivial ones. A system that recommends denial of disability benefits requires extensive documentation, while a system that sorts incoming mail does not. Relevant questions might include the following: Is the AI system advisory or determinative? Does a human review every output, or only flagged cases, or none? What discretion does the human reviewer have? Do they empirically use their discretion? The description should specify both what the system does and what it does not do, clarifying the human functions that remain even when AI assists.

*Data provenance.* What data was used to train the system, and what data does it process in operation? This element enables evaluation of whether the system's informational basis is appropriate for its task. A model trained on historical decisions may encode past biases, and a model processing incomplete records may produce unreliable outputs. Complete data documentation may sometimes be infeasible, but agencies should document what they know about where their data is from, if there are gaps or limitations in it, and the steps they have taken to address data quality issues.

*Performance evaluation.* How well does the system perform its intended function? This element requires agencies to specify metrics appropriate to whatever task the AI is aimed at, like accuracy, consistency, error rates, and processing speed, and to report measured performance against those metrics. Testing the system before deployment on relevant metrics would reduce the chance that harm occurs from a non-performant system, though some problems may only become apparent through use. Effective performance evaluation requires disaggregated analysis: does the system perform differently for different populations, case types, or contexts? If the system does perform differently across protected groups, that would create a red flag requiring explanation and, potentially, remediation, under the law.

---

[246] *See* Anthropic, *Responsible Scaling Policy* (Sept. 19, 2023, updated May 14, 2025), https://www.anthropic.com/rsp; OpenAI, *Preparedness Framework* (Dec. 18, 2023, updated Apr. 15, 2025), https://openai.com/index/updating-our-preparedness-framework/; Google DeepMind, *Frontier Safety Framework* (May 17, 2024, updated Sept. 22, 2025), https://storage.googleapis.com/deepmind-media/DeepMind.com/Blog/strengthening-our-frontier-safety-framework/frontier-safety-framework_3.pdf.

[247] Margaret Mitchell et al., *Model Cards for Model Reporting*, *in* FAT* '19: PROCEEDINGS OF THE CONFERENCE ON FAIRNESS, ACCOUNTABILITY, AND TRANSPARENCY 220, 220–29 (2019).

[248] Timnit Gebru et al., *Datasheets for Datasets*, 64 COMMC'NS ACM 86 (2021).

*Stress testing and red teaming.* How has the system been tested for failure modes? Adversarial testing can reveal vulnerabilities that ordinary performance evaluation misses, like inputs designed to fool the system, edge cases that the system struggles with, or scenarios where the system produces harmful outputs after a user or attacker manipulates it.[249] The science of red teaming advanced AI is improving, and should be integrated into agency evaluations of their systems before deployment. Agencies should document what testing was conducted, what vulnerabilities they identified in their systems, and what mitigation steps were taken. Where vulnerabilities were not remediable, the agency should explain the decision to deploy anyway and its response plan.

*Monitoring plan.* How will the system be monitored in operation? AI systems can drift over time as the world changes and the relationship between inputs and appropriate outputs shifts. For example, a model trained on pre-pandemic disability claims may perform poorly on post-pandemic cases involving long COVID, and people can respond to the use of AI by changing their behavior, making it shift off target. Agencies should specify how they will detect drift, what thresholds trigger review, and what retraining or adjustment procedures are in place. Consistent monitoring would be cheaper and less harmful than monitoring through lawsuits, where harm to a person is what causes the agency to check if its system is operating incorrectly.

*Explainability and traceability outputs.* What can be reconstructed about the reasoning behind individual decisions? This element does not require that AI systems be fully interpretable or even explainable, but it requires agencies to document what explanatory outputs are available. Feature importance scores, attention patterns, citation to evidence in the record, natural language rationales generated by the system and the like should be documented, along with known limitations of these explanatory outputs. As interpretability and similar sciences advance, their outputs could be integrated into this process.

---

[249] *See* Jessica Ji, *What Does AI Red-Teaming Actually Mean?*, CENTER FOR SECURITY & EMERGING TECHNOLOGY (Oct. 24, 2023), https://cset.georgetown.edu/article/what-does-ai-red-teaming-actually-mean/.

*Change log.* How has the system changed over time? Version control for AI systems would enable accountability as systems evolve. Was the model retrained? Was it fine-tuned? Did the data sources change? What changes were made to preprocessing or postprocessing? Why? The change log creates a historical record that courts and auditors can examine to understand how current system behavior relates to past documentation while not requiring full process at each update.

*Governance and accountability.* Who is responsible for the system? This element documents the organizational structure within the agency deploying the AI, including who owns it, who can authorize changes, who reviews performance, and what escalation pathways exist when problems arise. Responsible human officials can then be identified and held to account for any errors.

*Procurement and vendor constraints.* If the system was developed by an external vendor, what constraints govern? The Dossier should include audit rights, data access provisions, portability requirements, and any trade secret limitations on disclosure. Where vendor claims of trade secrecy limit this documentation,[250] agencies should explain what information is unavailable and why, enabling courts to evaluate whether the limitation is justified.

Of course, not every element of the Dossier requires exhaustive documentation for every system. Instead, it should be scaled to the stakes involved, requiring more extensive documentation for systems that make or influence high-stakes decisions and lighter documentation for lower-stakes applications. But the framework should be consistent across government, enabling cumulative learning about what AI documentation practices work and facilitating comparison across agencies and systems.

---

[250] *See* Note, *Wisconsin Supreme Court Requires Warning Before Use Of Algorithmic Risk Assessments In Sentencing*, 130 HARV. L. REV. 1530, 1530–37 (2017).

The Dossier would become part of the administrative record for any given agency action, subject to the same preservation and disclosure requirements as other parts of the record. It builds on existing OMB guidance around AI inventorying, monitoring, assessment, and procurement. But making it into a part of administrative law would give it a statutory basis that enables judicial review and enforcement. Litigants can contest agency action by identifying deficiencies in the Dossier, for example arguing that performance evaluation was inadequate, that known failure modes were not addressed, or that monitoring was insufficient, all according to set standards.

3. Material Model Change: When Updates Trigger Process

AI systems are rapidly developing, as the course of frontier AI over the past few years demonstrates. Advances in pre- and post-training, the advent of inference scaling, and continued algorithmic innovation have transformed how these systems operate and what they can do. Deployed systems can also change significantly over the course of deployment. These systems can be retrained as new data becomes available, adjusted as performance issues emerge, and updated as the underlying technology evolves. This consistent change creates a problem for administrative law, which assumes relatively discrete agency actions subject to defined procedural requirements based on relatively static foundations. When does an update to an AI system constitute "agency action" requiring notice and comment or other process?

The answer should balance accountability and the iterative development that AI enables. Requiring full process for every minor model adjustment would freeze AI deployment. But a change that significantly alters how the system behaves can be substantively equivalent to a change in the underlying rule, which would require real process. Allowing such changes without process would circumvent the procedural requirements that legitimate agency action.

As such, doctrine should be updated to track *material* model changes, that is updates that significantly affect outcomes, reasoning, or affected populations. Materiality is necessarily a standard rather than a rule, and its application can be guided by factors including the following:

*Outcome effects.* Does the update substantially change the rate at which something like disability benefit applications are approved or denied? Does it alter the distribution of outcomes across protected groups in ways that suggest a disparate impact? Changes that move outcome metrics by more than some threshold, defined in advance, would presumptively require enhanced process.

*Reasoning effects.* Does the update change which factors drive system outputs? A model that previously relied heavily on work history in its determinations might, after retraining, rely instead on medical evidence. In this kind of adjudication, even if aggregate outcomes are similar, the change in reasoning may affect how parties should present their cases and what evidence is relevant. To operationalize this kind of determination, technical interpretability tools could indicate if the activations in the system's neural network have changed when applied to a constant set of test cases before and after retraining to determine if a shift in reasoning has occurred.

*Population effects.* Does the update affect different populations differently? A change that improves overall accuracy while worsening performance for a particular demographic group could raise concerns about bias against that group.

*Architecture effects.* Does the update change the fundamental structure of the system rather than just updating parameters? Moving from one model architecture to another might warrant more scrutiny than retraining an existing architecture on new data.

When a material model change occurs, the agency should be required to update the Model and System Dossier, provide notice, and explain the expected impact of the change. This is less onerous than full notice-and-comment rulemaking but provides more notice than silent updates. The requirements should scale with the significance of the change.

The analogy to existing doctrine is the "logical outgrowth" test for final rules.[251] A final rule must be a logical outgrowth of an agency's proposed rule at the beginning of rulemaking. If the final rule differs so substantially that affected parties could not have anticipated it, the agency must provide a new opportunity for comment. Material model changes that alter the effective rule being applied trigger similar requirements, not because the formal rule has changed, but because the system implementing the rule operates differently. Congressional oversight of significant AI system changes might also provide a democratic check on *sub silentio* AI changes. This grounding also answers *Perez v. Mortgage Bankers Association*, which bars courts from demanding notice-and-comment for changes to interpretive rules beyond what the APA requires.[252] The material-model-change trigger keys to substance, not to updates as such. When a model change materially alters the operative standard, the agency has amended its rule in fact, and the APA's existing amendment process should apply.

4. Deference to Audit: A New Review Posture

How should courts review AI-assisted agency action? Not by deferring to substantive agency expertise and focusing on procedure, as they have traditionally done, nor by attempting to evaluate algorithms directly, but by assessing whether agencies have subjected their AI systems to rigorous technical audits. Auditing could help rebuild a foundation for substantive review of agency actions while putting them in a form that courts are able to handle, outsourcing technical complexity to technical experts but preserving accountability.

One traditional answer to how to do judicial review of agency AI action would be to require agencies to explain their AI-assisted decisions in terms that courts can evaluate, attempting to translate AI outputs into administrative law's reason-giving. This approach has surface appeal. It preserves familiar doctrinal forms and does not require courts to develop new technical competencies that they would struggle with. But it has serious problems. It invites *post hoc* rationalization, explanations for AI outputs that do not reflect the actual drivers of the decision (if those drivers are understood at all), and does not actually verify that the AI system is working properly but only that the agency can tell a good story about the system's outputs. These are also problems of pretextual human administrative action, but perhaps worsened in the AI case.

---

[251] *See* Patti Zettler, *The Logical Outgrowth Doctrine and FDA's Intended Use Revisions*, YALE J. REG. NOTICE & COMMENT (Jul. 18, 2017), https://www.yalejreg.com/nc/the-logical-outgrowth-doctrine-and-fdas-intended-use-revisions/.

[252] 575 U.S. 92 (2015).

A better approach shifts from deference to expertise toward deference to audit.[253] Under deference to audit, courts would still not seek to determine whether the agency's AI-assisted reasoning was substantively correct, because those questions would still be outside their expertise, but rather whether the agency subjected its AI systems to scrutiny and documentation. If an agency complies with a defined audit regime, for example by producing the Model and System Dossier, conducting performance evaluation and stress testing, maintaining monitoring and change logs, and getting independent verification that it had met auditing standards, courts would treat the technical components of the agency's action as presumptively reasonable. The agency here has earned deference not by claiming expertise but by demonstrating verification.

This safe harbor approach creates better incentives for agencies. Instead of current incentives, which point agencies toward defensive documentation in expectation of judicial review, agencies that want judicial approval for AI-assisted decisions would have a clear path, based on investing in auditing and documentation. Agencies that do not invest in auditing and verification would face more skeptical review and the burdens it brings. This doctrinal framework does not require courts to become machine learning experts, instead leveraging their procedural expertise to evaluate whether agencies have followed appropriate procedures for ensuring that their AI systems are trustworthy.

The safe harbor proposed here is not an absolute shield against challenge. Regulated parties and other claimants could still prevail against agencies by identifying specific flaws in agency AI systems, like evidence of bias that the agency's evaluations missed. The Dossier creates the evidentiary basis for such challenges, acting as both a shield for agencies that comply and a sword for challengers who can identify deficiencies. What challengers cannot do is simply assert that AI systems are untrustworthy without engaging the agency's documentation of system behavior and its efforts to ensure that the systems are acting reliably and well.

[253] The approach is kin to management-based regulation of AI, see Cary Coglianese, *Management-Based Oversight of the Automated State: Emerging Standards for AI Impact Assessment and Auditing in the Public Sector*, *in* GLOBAL PERSPECTIVES ON AI IMPACT ASSESSMENT (Emad Yaghmaei, et al., eds., forthcoming 2026).

This approach has antecedents in existing administrative law, expanding procedural review to the new AI context. Doctrinally, it keeps judicial review aimed at something like procedure, where courts are expert, while importing substantive review through technically sophisticated auditing. It also responds to the post-*Loper Bright* and post-*Slaughter* environment. The current Court is skeptical of agency claims to deference based on expertise. But deference to audit is different. It is deference to verifiable compliance with procedural requirements specified in law and checked by overseers. Courts and the legislature retain ultimate authority to define what audit requires (though it could also be delegated to internal functions of the Executive, for example based in OMB) and cede only the technical evaluation that they cannot perform themselves.

5. Institutional Architecture: Who Audits?

Deference to audit requires auditors. Courts can verify that agencies have produced documentation and followed specified procedures, but they cannot verify that the documentation is accurate or that the procedures that agencies followed are adequate. Three institutional mechanisms, operating in combination,[254] could help provide the audit capacity that this new era of AI accountability requires:

*Internal agency audit units.* Agencies that deploy AI systems should develop internal capacity to evaluate those systems. These agency AI audit units could implement the various requirements of the Model and System Dossier, setting requirements for agency records and determining whether they are met. However, internal audit is necessary but not sufficient. Auditors based within the agencies they are auditing may lack independence or share the blind spots of the teams they are intended to review.

---

[254] It might also be worth incentivizing private development of auditing capacity. *See* Gillian K. Hadfield & Jack Clark, *Regulatory Markets: The Future of AI Governance*, arXiv:2304.04914 (Apr. 25, 2023).

*External government auditors.* The Government Accountability Office (GAO) and agency Inspectors General could provide external audit capacity to supplement internal review. GAO already evaluates agency technology programs,[255] so extending this function to AI is a natural evolution. Inspectors General can investigate AI-related failures and assess whether internal audit processes are adequate. The US Center for AI Standards and Innovation (CAISI), sitting within NIST, already has some of the capabilities required for this kind of evaluation and standard-setting. CAISI or a similar body could answer questions about things like what a Model and System Dossier should contain for a specific function or what evaluations are necessary for specific systems. However, these bodies would need enhanced technical capacity, requiring investment.[256]

*Centralized review for high-impact systems.* Not all AI systems warrant the same level of scrutiny. A chatbot sitting on an agency website that provides personalized answers to frequently asked questions poses different risks than a system that makes welfare benefit denial determinations. A centralized review function, the AI analog to OIRA review of significant regulations, could focus intensive scrutiny on the AI systems that matter most. The threshold for centralized review should track the impact of the AI at issue. Systems that affect large numbers of people, that make or significantly influence high-stakes determinations, or that operate in sensitive domains, like law enforcement, immigration, and public benefits, would warrant more scrutiny than routine administrative applications or back-of-house automation. Exactly where this kind of capacity should sit is unclear, but it would need technical capacity. OIRA, or a new dedicated AI review function within OMB, could conduct this review. Such a central review body could also help identify common problems and promote solutions across the government, making overall uptake more efficient.

6. Remedies: What Courts Do When AI Governance Fails

[255] U.S. Gov't Accountability Office, *Science, Technology Assessment, and Analytics*, https://www.gao.gov/about/careers/our-teams/STAA.

[256] *See* West, *supra* note 239.

When AI-assisted agency action is unlawful, what remedies are appropriate? The standard remedial toolkit of vacatur, remand, and injunction[257] can be applied here, but AI systems present distinctive considerations that should guide their use.

*Vacatur versus remand.* Vacatur sets aside unlawful agency action while remand returns the matter to the agency for further proceedings. For AI-assisted adjudication at scale, vacatur may be impractical or harmful. Imagine an adjudication system in which an AI handles masses of claims. Setting aside all determinations made with such a system could affect millions of beneficiaries, many of whom might not have been harmed if the identified flaw did not relate to them, overwhelming agency capacity to redetermine cases and leaving claimants in limbo. Remand without vacatur may better balance remedial goals with practicability, though if the AI system systematically denied benefits to eligible claimants, vacatur may be necessary. On the other hand, if the flaw is procedural, remand to cure the deficiency may suffice.

*Targeted injunctions.* Injunctive relief can address specific failures in how an AI system has been developed or deployed without halting AI-assisted administration entirely. This is analogous to injunctions requiring specific procedural steps rather than enjoining agency action altogether.

*Disclosure under protective order.* Vendors sometimes claim that model specifications, training data, and performance evaluations are proprietary, and that their disclosure would harm competitive interests. Courts could respond through disclosure under protective order. Confidential submission to the court and, where appropriate, to challengers' experts, with restricted public dissemination, would avoid harm to the interests of companies and prevent trade secrecy from being an absolute bar to review.[258] If a vendor's system is so secret that no one can evaluate it, that system should not be making decisions that affect people's rights and undermines the democratic legitimacy of administration.

---

[257] 5 U.S.C. § 706(2) (2018).

[258] Fed. R. Civ. P. 26(c)(1).

*Structural remedies.* When the use of AI in administration creates systematic problems, reflecting inadequate agency capacity, vendor capture, or institutional dysfunction, case-by-case remedies may be insufficient. Courts could order structural relief, for example requiring external audits of agency uses of AI or procurement reforms.

### 7. Failure Modes and Hard Problems

No accountability regime is perfect. AI-assisted administration will fail in ways that even robust oversight cannot prevent, and some failures will cause harm in ways outside the protection of procedural mechanisms. Some such potential problems follow, as well as the beginnings of ways to address them, but more research is needed in this area.

*Hallucinated rationales.* AI systems, particularly ones based on LLMs, can generate plausible-sounding explanations that do not reflect the actual drivers of an output or decision.[259] A system that denies a benefit claim might produce an explanation citing factors that were not, in fact, determinative or that cites evidence that does not exist in the record. Agencies and courts may credit these explanations without recognizing their unreliability.

Mitigation requires grounding explanations in verifiable record materials. Explanations should cite specific evidence, citations should be checked against the actual record, and claims about what factors drove the decision should be tested against system behavior in similar cases. Technical interpretability research might also enable the evaluation of neuron activations inside LLMs to provide a substantive guide to their reasoning and what they considered. This process would not guarantee fidelity, but it reduces the risk of hallucination and creates opportunities for detection. Furthermore, hallucination rates have been steadily declining and should be compared to the rate of incorrect statements by humans in the relevant decision-making areas, which is probably higher than people expect.

---

[259] *See* Zhang et al., *supra* note 231.

*Automation bias.* Humans tend to defer to algorithmic recommendations in many domains, even if they are supposed to check whether the recommendation is appropriate.[260] In the administrative context, caseworkers reviewing AI-flagged cases might approve a system's recommendation without independently reasoning about the case or verifying if the AI was correct in its determination. If it happens across a large number of cases, such deference would undermine the extent to which human review is actually present, even when legally required.

Mitigation requires institutional design that would force human officials to actually use their judgment. Random audits of human responses to AI recommendations can be used to detect excessive agreement with AI recommendations, for example if a human overseer approves such recommendations at a pace that indicates that they cannot physically be checking the AI's determination. Override rates, determinations where the human rejects the AI recommendation, can also be monitored, and if a given official never overrides a system or seems to do so randomly, that would be evidence of no meaningful human oversight. Training can sensitize reviewers to automation bias. And for the highest-stakes decisions, procedures can require affirmative human justification with reasons given rather than simply affirming or denying AI outputs. Such proof of work echoes documentation requirements in traditional administrative law.

*Vendor capture.* Agencies may depend on vendors for access to frontier AI systems. Vendors may resist audit requirements that reveal system limitations, arguing that they must protect trade secrets, or design for lock-in, making it costly for agencies to switch providers. They may prioritize features that generate revenue over the public good.

The prescriptions here are familiar from other procurement contexts. Governments should require audit rights in contracts, portability requirements aimed to prevent lock-in, data access provisions that enable oversight by agencies, courts, and Congress. For core governmental functions, public development of AI systems to eliminate vendor dependence entirely may be necessary. The case for public AI infrastructure is strongest where the stakes for democracy, rights, welfare, and public accountability are highest. These domains should not depend on vendors whose interests may diverge from the public's where it is possible to avoid such dependence.

---

[260] *See* Jennifer M. Logg, Julia A. Minson & Don A. Moore, *Algorithm Appreciation: People Prefer Algorithmic to Human Judgment*, 151 ORGANIZATIONAL BEHAV. & HUM. DECISION PROCESSES 90 (2019).

*Value laundering.* AI systems may provide cover for value choices that should be made through democratic processes by laundering the underlying political question through opaque AI systems. Humans can blame machines to avoid taking blame themselves, point to alleged system error where a normative determination was the actual cause for the harm that people are objecting to. Relatively inscrutable AI systems are particularly useful for this kind of responsibility-avoidance. When used in this way, AI becomes a mechanism for laundering contestable value judgments through the appearance of technical neutrality.

As discussed above, the requirement that agencies identify normative choices that underpin algorithmic determinations is one principal safeguard against value laundering. When an AI system's behavior reflects value-laden parameters, like how much weight to place on different factors, what error rates to tolerate, or which tradeoffs to make, the agency must describe how those parameters were determined and tradeoffs handled in the administrative record.

### *D. Beyond Procedure: AI for Substantive Accountability*

The accountability mechanisms proposed thus far are mostly procedural in character, an odd turn in an Article partly about how procedure stifles government. They verify that agencies followed appropriate processes for developing and deploying AI systems without directly evaluating whether those systems reach correct conclusions. As I have argued, this is a familiar posture for administrative law, which has long substituted procedural for substantive review when the latter exceeds judicial competence.[261] And it invites an obvious objection: if AI accountability is just more procedure, will it not simply add to the ossification that already burdens administrative governance and prevent us from effectively adopting AI?

[261] *See supra* Part I.

The objection has force against a regime that relies on procedure alone. But there are three responses: First, AI will be integrated, if at all, through the existing frameworks and competencies of agencies, courts, and Congress. The framework above is built for that path. Second, the trap is, at bottom, a claim about cost curves: historically, the cost of overseeing administration has grown at least as fast as the complexity of what is overseen, so every capability gain came with an accompanying deficit that accumulated over time. The Fourth Settlement breaks the trap only if oversight cost grows sublinearly with system complexity. AI's distinctive promise is precisely that for the first time, that condition might obtain: verification is generally cheaper than generation; audits, once designed, standardize and automate across systems and cases; and the trust chain—of courts relying on auditors relying on evaluations—can itself be tested against outcomes in a way that "hard look" review of human reasoning can't. Importantly, this claim is falsifiable. If audit costs turn out to track model complexity linearly, the Dossier is another sediment layer and this Article's proposal fails. Deference to audit and adoption by agencies would let administrative law find out. Third, AI possesses characteristics that may enable forms of substantive accountability unavailable for human-staffed administration, potentially reducing the extent to which oversight must rely on ossifying procedure. Three areas deserve particular attention: technical alignment, interpretability, and scalable oversight. Together, they suggest that AI-assisted governance can be subject to verification mechanisms that go beyond process to examine whether systems are actually pursuing appropriate objectives through appropriate means. Though the technology is nascent and its course speculative, to the extent these mechanisms mature and prove reliable, they reduce rather than increase the need for procedural safeguards and so could help de-ossify government.

*Technical alignment.* Human employees of administrative agencies have their own goals, like career advancement, ideological commitments, cognitive ease, and personal relationships, that only partially overlap with their agencies' statutory missions.[262] Organizations attempt to align employee behavior through compensation, promotion, discipline, professional socialization, and hierarchical oversight, but these mechanisms are costly and imperfect.[263] They shape behavior at the margin but cannot ensure that employees genuinely internalize organizational objectives. The principal-agent problems that pervade bureaucracy reflect the fundamental difficulty of aligning human goals to institutional purposes.[264]

[262] Avinash Dixit, *Incentives and Organizations in the Public Sector: An Interpretative Review*, 37 J. HUM. RES. 696, 696–700, 708–12 (2002).

[263] *Id.*

AI systems can be aligned through different mechanisms. Reinforcement learning from human feedback fine-tunes model outputs to match human evaluators' preferences.[265] Constitutional AI extends this approach to AI graders, making it more efficient and effective,[266] while deliberative alignment extends it to AI "reasoning."[267] Reward modeling and preference learning shape system behavior toward specified goals. These techniques, and their cousins, are still nascent and face problems like reward hacking,[268] goal misgeneralization,[269] and alignment faking.[270] But they operate on the system's objectives more directly than the external incentives used to align humans. A technically aligned AI system pursues its specified objectives because its training has in some sense made those objectives its own, not because external rewards and punishments make compliance instrumentally rational, representing a real improvement on human alignment.

The implications for administrative accountability are significant. To the extent an AI system is reliably aligned to statutory objectives, the oversight burden shifts. We need not verify through procedural proxies that the system is trying to achieve the right goals or even check its work substantively because we can have reasonable confidence in its objectives and focus oversight on whether it is achieving them effectively. Alignment does not eliminate accountability—systems can be aligned to the wrong objectives or can fail to achieve objectives they are aligned to—but it changes what accountability through oversight must accomplish. Procedural requirements designed to detect goal divergence substitute for the alignment we cannot achieve in humans, and to the extent alignment can be technically verified, that substitution can shrink.

[264] *See* Jensen & Meckling, *supra* note 204 at 308–10.

[265] Ouyang et al., *supra* note 26.

[266] Bai et al., *supra* note 26.

[267] Melody Y. Guan et al., *Deliberative Alignment: Reasoning Enables Safer Language Models*, arXiv:2412.16339 (Dec. 20, 2024).

[268] Dario Amodei et al., *Concrete Problems in AI Safety*, arXiv:1606.06565 (Jul. 25, 2016).

[269] Lauro Langosco et al., *Goal Misgeneralization in Deep Reinforcement Learning*, 162 PMLR 12004 (2022).

[270] Ryan Greenblatt et al., *Alignment faking in large language models*, arXiv:2412.14093 (Dec. 18, 2024).

*Interpretability.* Human reasoning is opaque. We can ask an expert why she reached a conclusion, but her introspective report may be inaccurate, incomplete, or strategically constructed. Cognitive science has demonstrated that humans often cannot accurately report the factors that influenced their judgments and that *post hoc* rationalization is the norm rather than the exception.[271] This has serious implications for administration, which relies upon delegating important decisions to opaque human agents. When a claims examiner denies a disability application, we cannot verify that the stated reasons reflect the actual cognitive process that produced the denial. The examiner herself may not know.

This opacity is a primary driver of procedural substitution. If we cannot examine the substance of expert reasoning, we demand that experts document their reasoning in reviewable form, like the records, explanations, and responses to objections that make up so much administrative paperwork. This documentation requirement creates accountability, but it is accountability for what experts say they did, not for what they actually did. Sophisticated actors learn to produce documentation that survives review regardless of the underlying reasoning process.

AI systems are not introspectively transparent, but they are mechanistically inspectable in ways that human brains are not. Interpretability research aims to understand how neural networks process information and produce outputs, including which features activate in response to which inputs, how information flows through network layers, and what internal representations the system constructs.[272] The field of interpretability is young, and current techniques provide limited insight into large model behavior. But to the extent that interpretability works, and it is developing over time, it could enable a form of substantive review that human oversight of human experts cannot achieve. Human minds are true black boxes, a problem that administrative law has handled by proxy through procedural reasoning requirements. But with interpretability, overseers could in theory examine an AI system's actual reasoning process. For example, interpretability might show an activation pattern associated with deception or pretextuality, or it might show activations that indicate that a model was actually reasoning about an alternative proposed in a rulemaking comment.

---

[271] Richard E. Nisbett & Timothy D. Wilson, *Telling More Than We Can Know: Verbal Reports on Mental Processes*, 84 PSYCHOL. REV. 231 (1977).

[272] *See* Santhosh Kumar Ravindran, *Adversarial Activation Patching: A Framework for Detecting and Mitigating Emergent Deception in Safety-Aligned Transformers*, arXiv:2507.09406 (Jul. 12, 2025).

Interpretability-based review is not a magic wand, and it would not replace traditional accountability mechanisms entirely. Interpretability techniques require technical expertise to apply, and there is ongoing debate about when interpretations of these black box models are meaningful.[273] But interpretability outputs could be integrated into the auditing and record processes that ground judicial review. Where interpretability works, it could occasion a significant advance in how review can occur. Courts could verify that appropriate interpretability analysis was conducted and that technical review of that analysis occurred, much as they currently verify that appropriate scientific analysis was conducted and evaluate agency reliance on it.

In the limit, the combination of alignment and interpretability suggests the possibility of a different accountability posture than pure proceduralism. For AI systems that are both technically aligned to statutory objectives through validated alignment techniques and subject to interpretability analysis verifying that their reasoning processes track legal requirements, substantive accountability becomes tractable in ways it is not for human administrators. The procedural requirements that have come into being since 1887 because substantive review became impossible could be relaxed when substantive review becomes more possible. As such, the burdens of paperwork could be lifted as the system produces more accountable government.

*Scalable oversight.* Oversight could also begin to include AI systems that oversee other AI systems. Scalable oversight and control mechanisms are being developed in frontier AI developers and could be applied in the governmental context. Human oversight of administrative action faces the fundamental cognitive bottlenecks described in Part III.A above. The millions of individual determinations that agencies make annually cannot all receive human review.[274] Sampling, prioritization, and triage are necessary compromises, but most agency actions escape external scrutiny entirely because there is simply not enough human cognition to go around.

[273] *See* Cynthia Rudin, *Stop explaining black box machine learning models for high stakes decisions and use interpretable models instead*, 1 NATURE MACHINE INTELLIGENCE 206 (2019); Gillian K. Hadfield, *Explanation and justification: AI decision-making, law, and the rights of citizens*, SCHWARTZ REISMAN INSTITUTE FOR TECHNOLOGY AND SOCIETY (May 18, 2021), https://srinstitute.utoronto.ca/news/hadfield-justifiable-ai.

AI oversight does not face the same constraint. Spinning up a new AI instance is cheap. A system designed to audit other systems could in theory process every case where that AI makes a decision, not just a sample, especially if the costs of auditing are less than making the initial decision, as they often likely would be. This overseer AI could identify cases and patterns warranting human review, compare decisions across adjudicators to detect inconsistency, monitor outcomes over time to detect drift, and examine reasoning processes to detect deviation from expected patterns. We might prefer human to AI oversight in many cases for normative reasons, but the alternative here is often not human review but no review, and AI oversight would extend accountability to actions that might otherwise escape review entirely. Furthermore, human attention then could be concentrated on the hard or normatively important Type 3 cases that AI surfaces, or on setting overall rules and breaking normative ties, providing broader coverage than human review alone while preserving human authority over consequential judgments.

*Implications for ossification.* If alignment, interpretability, and scalable oversight mature as accountability mechanisms, they would change the calculus of accountability and procedural design. If AI systems can be aligned to appropriate objectives, if their reasoning processes can be examined through interpretability, and if their outputs can be monitored comprehensively through scalable oversight, the need for procedural proxies would diminish. As such, the reliance on procedure that currently characterizes administrative law could decline over time, another dimension along which AI would improve government function.

Even in the bull case for these technical improvements, procedure would still be necessary. Process requirements serve multiple functions beyond enabling substantive accountability, including providing participation opportunities, creating deliberation-forcing mechanisms, and establishing legitimacy through public engagement. These functions would persist even if substantive accountability becomes more tractable, and might be even more necessary as government is increasingly run through AI. But the accretive proceduralism that characterizes modern administrative law, where each oversight failure generates new procedural requirements layered atop existing ones, can be mitigated if substantive verification becomes possible.

[274] Bowman et al., *supra* note 28.

The anti-ossification potential of AI accountability is thus twofold. First, the mechanisms proposed in Part III.C above, including the Dossier, audit-based review, and centralized oversight, would be standardized and applicable across cases, reducing the procedural burden of individual processing that slows agency action and increasing the cognitive capacity available to handle it. Second, to the extent alignment and interpretability enable substantive accountability, they reduce the need for procedures that exist only as substitutes for substance. Ideally, this potential engenders a rebalancing toward the substantive oversight of administration that procedure was always a second-best replacement for.

Of course, this rebalancing is contingent on technical progress that remains uncertain. Alignment techniques may plateau short of reliable goal verification, interpretability may never fully illuminate large model reasoning and behavior, and scalable oversight may introduce its own errors and biases. But if it works, even in limited forms, it could improve substantive oversight because AI possesses characteristics enabling forms of verification that human-staffed administration does not. Administrative law should be designed to take advantage of these characteristics as they mature, creating incentives for agencies to invest in alignment and interpretability rather than treating procedure as the only path to legitimacy.

## CONCLUSION: CRISIS AS OPPORTUNITY

The history of administrative law is defined by technological crisis and technological opportunity. From the railroads and the advent of modern administration in the ICC, to the Great Depression, the New Deal agencies created to handle it, and the APA, to the environmental crises that precipitated the integration of complex science and computerization into administration and the *Chevron* settlement, crises have led to disruption and administrative resettlement.

This cycle has generally been defined by crises of capability and accountability. Government has adopted new technology to handle new problems, and its enhanced capacities have led to overreach that creates backlash aiming to restore accountability. But the present moment is anomalous. The Supreme Court and the President together are retrenching the administrative state without a precipitating capability crisis. Today, the crisis is not governmental overreach but governmental inscrutability. From *Loper Bright* to *Slaughter*, the doctrinal shifts underway are a scrutability intervention that risks sacrificing capability for comprehensibility at precisely the moment when complex new problems demand sophisticated responses. Climate change, pandemics, and advanced AI present problems for which existing regulatory frameworks are unprepared, and these problems will not wait for government to resolve its internal contradictions. They will force resolution.

History suggests that moments of crisis create openings for institutional transformation that are otherwise foreclosed. The procedural sediment of administrative burdens described in Part II accumulated precisely because, absent crisis, incumbent interests can block reforms that threaten their bargaining positions. In crises, the political barriers to transformation temporarily fall. And new technologies present a novel opportunity for administrative transformation, if we are able to seize it.

This Article has argued that AI offers tools for a new kind of administration and a new kind of settlement in administrative law. These tools could operate as scrutability infrastructure that lowers the cognitive costs of participation and oversight rather than raising them, making government more responsive and more effective than ever before. The law should combine with technology to build a new administrative law framework for AI that opens its transformative potential.

A more fundamental contribution might lie in AI's potential to restore substantive accountability, reversing the course toward proceduralization that has been in progress since the beginning of the twentieth century. The goal is verifying that government is actually pursuing appropriate objectives through appropriate means, not merely checking whether it has followed required procedures. Technical alignment, interpretability, and scalable oversight offer mechanisms for such verification that do not exist for human-staffed administration. If these capabilities mature, and their trajectory, while uncertain, is promising, they could enable a settlement that previous technologies could not, one based on verified substance and with reduced reliance on procedure.

The capability-accountability trap has structured American government and its administrative law for over a century. It need not structure the next century. The current administrative retrenchment and our oncoming problems present an opportunity for change. If the reformers who built previous settlements could match the railroad with the expert commission, the Depression with the APA, and complex science with the framework of deferential review, this generation can match the age of artificial intelligence with administrative law suited to its challenges. This Article has laid out the beginnings of a path to that kind of administrative law.